\documentclass[aps,prd,showpacs,amssymb,floatfix,nofootinbib,superscriptaddress]{revtex4-1}
\usepackage{graphicx,amsmath,subfigure,psfrag}
\usepackage{multirow} 
\usepackage{xspace}
\usepackage{bm}
\usepackage{enumerate}

\newcommand{\eeq}{\end{equation}}
\newcommand{\bea}{\begin{eqnarray}}

\def\ltsima{$\; \buildrel < \over \sim \;$}
\def\simlt{\lower.5ex\hbox{\ltsima}}
\def\gtsima{$\; \buildrel > \over \sim \;$}
\def\simgt{\lower.5ex\hbox{\gtsima}}

\newcommand{\OrbitalPhase}{\Phi}

\newcommand{\OrbitalFreq}{\Omega}

\newcommand{\abs}[1]{\left\lvert #1 \right\rvert}
\newcommand{\define}{\equiv}

\def\lesssim{\mathrel{\hbox{\rlap{\hbox{\lower4pt\hbox{$\sim$}}}\hbox{$<$}}}}
\def\gtrsim{\mathrel{\hbox{\rlap{\hbox{\lower4pt\hbox{$\sim$}}}\hbox{$>$}}}}
\def\alt{\mathrel{\hbox{\rlap{\hbox{\lower4pt\hbox{$\sim$}}}\hbox{$<$}}}}
\def\agt{\mathrel{\hbox{\rlap{\hbox{\lower4pt\hbox{$\sim$}}}\hbox{$>$}}}}
\def\PRD{{\it Phys. Rev.} D~}
\def\PRL{{\it Phys.Rev.} Lett~}

\def\CQG{{\it Class. Quantum Grav.}}

\def\gta{\ifmmode {\mathbin{\lower 3pt\hbox   
    {$\,\rlap{\raise 5pt\hbox{$\char'076$}}\mathchar"7218\,$}}}
    \else {${\mathbin{\lower 3pt\hbox
    {$\rlap{\raise 5pt\hbox{$\char'076$}}\mathchar"7218\,$}}}
    $}\fi}
\def\lta{\ifmmode {\,\mathbin{\lower 3pt\hbox   
    {$\,\rlap{\raise 5pt\hbox{$\char'074$}}\mathchar"7218\,$}}}
    \else {${\mathbin{\lower 3pt\hbox
    {$\rlap{\raise 5pt\hbox{$\char'074$}}\mathchar"7218\,$}}}
    $}\fi}

\newcommand{\SU}{\affiliation{Department of Physics, Syracuse University, Syracuse, NY 13244, USA.}}
\newcommand{\CU}{\affiliation{Institute of Astronomy, Madingley Road, CB3 0HA Cambridge, UK.}}

\begin{document}
\title{Accurate modeling of intermediate-mass-ratio inspirals: Exploring the form of the self-force in the intermediate-mass-ratio regime}

\author{E.A. Huerta}\email{eahuerta@syr.edu}\SU\CU%
\author{Prayush Kumar}\email{prkumar@syr.edu}\SU%
\author{ Duncan A. Brown}\email{dabrown@phy.syr.edu}\SU%


\date{\today}

\begin{abstract}        
In this paper we develop a waveform model that accurately reproduces the dynamical evolution of intermediate-mass-ratio inspirals, as predicted by the effective-one-body (EOB) model introduced in \cite{buho}, and which enables us to shed some light on the form of the self-force for events with mass-ratio 1:6, 1:10 and 1:100. To complement this study, we make use of self-force results in the extreme-mass-ratio regime, and of predictions of the EOB model introduced in \cite{buho}, to derive a prescription for the shift of the orbital frequency at the innermost stable circular orbit which consistently captures predictions from the extreme, intermediate and comparable mass-ratio regimes.

\end{abstract}

\pacs{}

\maketitle

\section{Introduction}    

Testing Einstein's theory of general relativity in the strong-field regime by directly detecting the gravitational waves (GWs) emitted by black hole (BH) binaries of extreme mass-ratio has revived interest in a fundamental problem in general relativity: that of the gravitational self-force acting on a mass particle that moves in the background of a more massive BH. The gravitational self-force arises as  a result of the back reaction between the  small compact object and its own gravitational field, which, at linear order in mass-ratio, corresponds to a linear perturbation of the central BH geometry. Theoretical work in this field has been making steady progress since the seminal contributions of Dirac on the electromagnetic self-force in flat spacetime \cite{dirac}, and the extension of this analysis to curved spacetime by DeWitt an Brehme \cite{dewitt}.   The generalization of these early studies to the case of the gravitational self-force was accomplished independently by Quinn and Wald \cite{qwald} and Mino, Sasaki and Tanaka \cite{mino}. More recent developments have introduced mathematical rigor in the theoretical derivation of the gravitational self-force, and have relaxed previous assumptions with regard to the internal structure of the small compact object, i.e, it can now be a small Kerr BH or a small compact object made up of ordinary matter \cite{grallaI,grallaII}.

Likewise, the actual computation of the gravitational self-force has evolved from  simplified scalar-field models~\cite{scasf}, to more sophisticated solutions that involve electromagnetic and gravitational problems in the context  of Schwarzschild circular orbits. At present, the self-force program has succeeded in developing numerical codes to compute the gravitational self-force along generic orbits around a Schwarzschild BH, and actually implementing these computations to develop an accurate waveform model that describes the inspiral evolution of non-spinning stellar mass BHs onto supermassive non-spinning BHs \cite{wargar}. The development of numerical algorithms to compute the self-force on a scalar charge moving along an eccentric-equatorial orbit of a Kerr BH has also been accomplished \cite{war}. The extension of this algorithm to compute the gravitational self-force for Kerr inspirals is under development  \cite{war,warleor}. 

The fact that the self-force program has focused on the computation of the self-force for spinless particles that inspiral into more massive BHs has a physical rationale. It is not just that such a problem would be more difficult to solve. It has also been shown that in the context of extreme-mass ratio inspirals (EMRIs), with typical mass-ratios 1:\(10^{6}\), the inclusion of small-body spin corrections in search templates will not allow us to measure the small body spin parameter with good accuracy  \cite{smallbody}. Hence, from a data analysis perspective, the inclusion of small body spin effects is not necessary.  Additionally, before attempting to compute the self-force for spinning particles, one may need to address a more pressing modeling issue for spinless particles: it has been shown that including conservative self-force corrections in the orbital phase of EMRIs may not be necessary for source detection, but they may still be necessary for accurate parameter reconstruction. Additionally, second order radiative corrections may contribute to the phase evolution at the same level as first-order conservative corrections \cite{cons,conspro}.  This then suggests that a waveform template that aims to provide an accurate description of the inspiral evolution of EMRIs may have to include both first-order conservative corrections, and second-order radiative corrections, as pointed out in \cite{wargar}.

In sharp contrast, modeling BH binaries with intermediate mass ratio, i.e., 1:10-1:1000, (IMRIs) presents new challenges that can be neglected in the EMRI arena.  In the absence of fully general relativistic gravitational self-force corrections in this mass-ratio regime, some studies have assessed the importance of including post-Newtonian self-force corrections in search templates for spinning BHs of intermediate-mass that inspiral into supermassive Kerr BHs. This work has shown that the implementation of first-order post-Newtonian self-force corrections for spin-spin and spin-orbit couplings is essential to ensure the reliability of parameter estimation results \cite{higherspin}.  These studies suggest that the computation and implementation of gravitational self-force corrections for spinning binaries is a pressing, important problem both from a theoretical and data analysis perspective.

In this article we shed light on the importance of including gravitational self-force corrections in search templates for non-spinning stellar mass BHs that inspiral into Schwarzschild BHs of intermediate-mass. As, at present, we have no access to accurate self-force calculations in this mass-ratio regime, it is important to shed light on the regime in which current self-force corrections do not render an accurate dynamical evolution of GW sources, and explore the range of applicability of the information we have at hand to develop accurate waveform templates in that mass-ratio regime. This study is particularly important in view of the ongoing upgrade of the LIGO detector~\cite{aLIGO}. Once advanced LIGO (aLIGO) begins observations, it will be possible to target the inspirals of neutron stars (NSs) and stellar-mass BHs into intermediate-mass BHs with masses  \(\sim 50M_{\odot} - 350M_{\odot}\)~\cite{brown} ---events which may take place in core-collapsed globular clusters~\cite{evidence,firstpaper}.  To perform this study we will make use of the recently developed effective-one-body (EOB) model that has been calibrated using numerical relativity (NR)  simulations for non-spinning BH binaries of mass-ratio    \(q=m_1/m_2\) \(=1,2,3,4\) and 6 \cite{buho}. It is worth pointing out that the actual calibration of this EOBNRv2 (EOBNR version two) model reproduces with great accuracy the features of true inspirals, and hence we will use this model as a benchmark to explore the form of the self-force in the intermediate-mass ratio regime. By construction, the EOBNRv2 model reproduces results in the test-mass particle limit, and also encodes self-force corrections that have been derived for small mass-ratios. In our analysis we will explicitly show these important modeling ingredients when we compare the predictions made by the EOBNRv2 and those obtained through black hole perturbation theory (BHPT).  

Finally, we will make use of the EOBNRv2 model and pertubative results to present a new prescription for the orbital frequency shift at the innermost stable circular orbit (ISCO), originally derived in \cite{inner} in the context of EMRIs. Our prescription reproduces exactly the self-force prediction for extreme-mass ratios and provides an accurate prediction in the intermediate and comparable-mass ratio regimes.   

This paper is organized as follows. In Section~\ref{s1} we present a succinct description of the EOBNRv2 model. In Section~\ref{s2} we derive an IMRI waveform model that accurately captures the features of true inspirals, as compared with the EOBNRv2 model. The derivation of this model will enable us to explore what the form of the self-force should be in the intermediate-mass-ratio regime so as to accurately reproduce the orbital dynamics obtained through NR simulations. In Section~\ref{s3} we derive a new prescription for the gravitational self-force correction to the orbital frequency at the ISCO that encodes results from the extreme, intermediate and comparable-mass ratio regimes. Finally, we summarize our results in Section~\ref{s4}.

\section{Effective One Body model}
\label{s1}

The Effective One Body (EOB) model was recently calibrated for non-spinning BH binaries of mass ratios   \(q=m_1/m_2\) \(=1,2,3,4\) and 6 by comparison to NR simulations \cite{buho}. In this Section we briefly describe this model.

\subsection{EOB dynamics}

The EOB is a scheme that maps the dynamics of the two body problem in general relativity to that of one object moving in the background of an effective metric. In the non-spinning limit this metric takes the form 

\begin{equation}
  ds_\mathrm{eff}^2 = -A(r)\,dt^2 + \frac{D(r)}{A(r)}\,dr^2 +
  r^2\,\Big(d\Theta^2+\sin^2\Theta\,d\OrbitalPhase^2\Big) \,,
  \label{eq:EOBmetric}
\end{equation}

\noindent where  $(r,\OrbitalPhase)$ stand for the dimensionless radial and polar coordinates, respectively. The conjugate momenta of these quantities is given by $(p_r,p_\OrbitalPhase)$. Since \(p_r\) diverges near the horizon, it is convenient to replace it  by the momentum conjugate to the EOB \textit{tortoise} radial coordinate $r_*$, i.e., 

\begin{equation}
  \frac{dr_*}{dr}=\frac{\sqrt{D(r)}}{A(r)}\,.
\end{equation}

\noindent Using this coordinate transformation, the effective EOB Hamiltonian can be written as  \cite{buho}

\begin{equation}
  \label{eq:genexp}
  H^\mathrm{eff}(r,p_{r_*},p_\OrbitalPhase) \define \mu\,\widehat{H}^\mathrm{eff}(r,p_{r_*},p_\Phi)  \\
  = \mu\,\sqrt{p^2_{r_*}+A (r) \left[ 1 +
      \frac{p_\OrbitalPhase^2}{r^2} +
      2(4-3\eta)\,\eta\,\frac{p_{r_*}^4}{r^2} \right]} \,,
\end{equation}

\noindent where \(\mu = m_1 m_2/(m_1+m_2)\), \(M=m_1+m_2\) and \(\eta=\mu/M\) stand for the reduced and total mass of the system, and the symmetric-mass ratio, respectively.  Additionally, the real EOB Hamiltonian is given by \cite{buho}

\begin{equation}
  \label{himpr}
  H^\mathrm{real}(r,p_{r_*},p_\OrbitalPhase) \define \mu\hat{H}^\mathrm{real}(r,p_{r_*},p_\Phi)  \\
  = M\,\sqrt{1 + 2\eta\,\left ( \frac{H^\mathrm{eff} - \mu}{\mu}\right )}
  -M\,.
\end{equation}

\noindent Furthermore, to ensure the existence and \(\eta\)-continuity of a last stable orbit (ISCO) as well as the existence and \(\eta\)-continuity of an \(\eta\)-deformed analog of the light-ring (the last stable orbit of a massless particle), these metric coefficients must be Pad\'e resummed. At present these coefficients are available at (pseudo) 5PN and 3PN order for  \(A(r)\) and \(D(r)\), respectively,

\begin{equation}
  D(r)=\frac{r^{3}} {(52\,\eta - 6\,\eta^{2}) + 6\, \eta\, r +
    r^{3}}\,,
\end{equation}

\noindent and 

\begin{equation}
  A(r) = \frac{\mathrm{Num}(A)}{\mathrm{Den}(A)}\,,
\end{equation}
with
\begin{eqnarray} \mathrm{Num}(A) &=& r^4\,\left[-64 +
    12\,a_4+4\,a_5+a_6+64 \eta-4 \eta ^2 \right]
  \nonumber\\
  &+& r^5\,\left[32-4\,a_4-a_5-24 \eta \right]\,,
\end{eqnarray}
and
\begin{eqnarray}
  && \mathrm{Den}(A) = 4\,a_4^2+4\,a_4\,a_5+a_5^2-a_4\,a_6+16\,a_6+ (32\,a_4 \nonumber \\
  &&\qquad + 16\,a_5-8\,a_6)\,\eta + 4\,a_4\,\eta^2+32\,\eta^3 + r\,\left[4\,a_4^2+a_4\,a_5 \right. 
  \nonumber\\
  &&\qquad \left. +16\,a_5+8\,a_6+(32\,a_4 -2\,a_6)\,\eta + 32\,\eta^2+8\,\eta^3\right] 
  \nonumber\\
  &&\qquad + r^2\,\left[16\,a_4+8\,a_5+4\,a_6+(8\,a_4+2\,a_5)\,\eta +32\,\eta^2\right] \nonumber \\
  &&\qquad + r^3\,\left[8\,a_4+4\,a_5+2\,a_6+32\,\eta-8\,\eta^2\right] 
  \nonumber\\
  &&\qquad + r^4\,\left[4\,a_4+2\,a_5+a_6+16\,\eta-4\,\eta^2\right] \nonumber \\
  &&\qquad + r^5\,\left[32-4\,a_4-a_5-24\,\eta\right]\,,
\end{eqnarray}
\noindent where $a_4=[94/3-(41/32)\,\pi^2]\,\eta$, and \(a_5\), \(a_6\) are adjustable parameters which were determined by minimizing the inspiral phase difference between the NR and EOB (2,2) modes in \cite{buho}. 

The EOB equations of motion that describe the orbital dynamics of the BH binary are given by \cite{bur}

\begin{subequations} \label{eq-eob}
  \begin{align}
    \frac{dr}{d \widehat{t}} &=
    \frac{A(r)}{\sqrt{D(r)}}\frac{\partial
      \widehat{H}^\mathrm{real}} {\partial
      p_{r_*}}(r,p_{r_*},p_\OrbitalPhase)\,,
    \label{eq:eobhamone} \\
    \frac{d \OrbitalPhase}{d \widehat{t}} &= \frac{\partial
      \widehat{H}^\mathrm{real}} {\partial
      p_\OrbitalPhase}(r,p_{r_*},p_\OrbitalPhase)\,,
    \label{eq:eobhamtwo}\\
    \frac{d p_{r_*}}{d \widehat{t}} &=
    -\frac{A(r)}{\sqrt{D(r)}}\,\frac{\partial \widehat{H}^\mathrm{
        real}} {\partial r}(r,p_{r_*},p_\OrbitalPhase) +{}^\mathrm{
      nK}\widehat{\cal F}_\OrbitalPhase \,
    \frac{p_{r_*}}{p_\OrbitalPhase}\,, \label{eq:eobhamthree}\\
    \frac{d p_\OrbitalPhase}{d \widehat{t}} &= {}^\mathrm{
      nK}\widehat{\cal F}_\OrbitalPhase\,,
    \label{eq:eobhamfour}
  \end{align}
\end{subequations}

\noindent where $\hat{t}\equiv t/M$, $\widehat{\OrbitalFreq}\define d \OrbitalPhase/d \widehat{t} \define M\Omega$ and the radiation-reaction force \({}^\mathrm{nK}\widehat{\cal F}_\OrbitalPhase \) is given by \cite{buho}
 
\begin{equation}\label{RadReacForce} {}^\mathrm{nK}\widehat{\cal
    F}_\OrbitalPhase = -\frac{1}{\eta\,v_\Omega^3}\,\frac{dE}{dt}\,,
\end{equation}

\noindent with $v_\Omega \define \widehat{\OrbitalFreq}^{1/3}$. An important improvement in the EOB formalism over models that used Pad\'e resummation of Taylor approximants to the energy flux is the implementation of a resummed energy flux of the form

 \begin{equation}\label{resflux}
  \frac{dE}{dt}=\frac{v_\Omega^6}{8\pi}\,
  \sum_{\ell=2}^{8}\,\sum_{m=1}^{\ell}\, m^2\, \abs{
    \frac{\mathcal{R}}{M}\, h_{\ell m}}^2~,
\end{equation}

\noindent where \( \mathcal{R}\) is the distance to the source and $h_{lm}$'s represent the multipoles of the waveform, defined through the following relation 

\begin{equation}
\label{mulwav}
h_{+} - i h_{\times} = \frac{M}{\mathcal{R}} \sum^{\infty}_{l=2} \sum^{m=l}_{m = -l} Y^{lm}_{-2}\, h_{lm},
\end{equation}

\noindent where $Y^{lm}_{-2}$ represent the spin weighted -2 spherical harmonics and  $h_+$ and $h_{\times}$ stand for the two gravitational wave polarizations. Since the numerical relativity simulations used to calibrate the EOB model were found to satisfy the condition $h_{\ell  m}=(-1)^{\ell}\,h_{\ell\,-m}^*$ with great accuracy, where $^*$ denotes complex conjugate, one can also assume that the analytical modes that enter the sum in Eq.~\ref{resflux} also satisfy this property. Furthermore, because $\abs{h_{\ell, m}}=\abs{h_{\ell,-m}}$, the sum in  Eq.~\ref{resflux} extends only over positive \(m\) modes.

\clearpage 

\subsection{Modelling of the inspiral and plunge evolution}  

In the EOB scheme, the inspiral and plunge phases are described by the product of several factors, namely, 

\begin{equation}\label{hip}
  h^\mathrm{insp-plunge}_{\ell m} = h^\mathrm{F}_{\ell m}\,N_{\ell m}\,,
\end{equation}

 \noindent where the function \(N_{\ell m}\) is introduced to ensure that the EOB model reproduces: a) the shape of the NR amplitudes \(|h_{\ell m}|\) near their maxima; and b) the timelag between the maxima of the \(|h_{\ell m}|\) and the maxima of \(|h_{22}|\), obtained from NR data. On the other hand, the factorized resummed modes \(h^\mathrm{F}_{\ell m}\) are given by
 
 \begin{equation}\label{hlm}
  h^\mathrm{F}_{\ell m}=h_{\ell m}^{(N,\epsilon)}\,\hat{S}_\mathrm{eff}^{(\epsilon)}\, T_{\ell m}\, e^{i\delta_{\ell m}}\,\left(\rho_{\ell m}\right)^\ell\,,
\end{equation}

\noindent where \(\epsilon\) stands for the parity of \(h_{\ell m}\), i.e., \(\epsilon\,=\,1\) if \( \ell+m\) is even, and \(\epsilon\,=\,0\) for odd \( \ell+m\).

The factor  \(h_{\ell m}^{(N,\epsilon)}\) stands for the Newtonian contribution, defined in Eqs.(15)-(18) of \cite{buho}. The remaining terms \(\hat{h}_{\ell m}^{(\epsilon)}= \hat{S}_\mathrm{eff}^{(\epsilon)}\, T_{\ell m}\, e^{i\delta_{\ell m}}\left(\rho_{\ell m}\right)^\ell\) represent a resummed version of all PN corrections, which have the structure \(\hat{h}_{\ell m}^{(\epsilon)}= 1+ {\mathcal{O}}(x)\), where \(x\) is the gauge-invariant object \(x=\hat{\Omega}^{2/3}\).

Regarding the structure of the \(\hat{S}_\mathrm{eff}^{(\epsilon)}\) factor, we note that in the even parity case, which corresponds to mass moments, the leading order source of GW radiation is given by the energy density. Therefore, the source factor can be defined as  \(\hat{S}_\mathrm{eff}^{(\epsilon=0)} =  \hat{H}^\mathrm{eff}(r, p_{r_*}, p_\Phi)\)  \cite{resu}. On the other hand,  in the odd-parity case, which is associated to current modes, the angular momentum \(\hat{L}_\mathrm{eff}\) turns out to be a factor in the Regee-Wheeler-Zerilli odd-parity multipoles in the limit of small mass-ratio \(\eta\) \cite{resu}. Hence, one can define \(\hat{S}_\mathrm{eff}^{(\epsilon=1)} =  \hat{L}_\mathrm{eff}= p_\Phi\, v_\Omega\)  \cite{buho}.

Furthermore, considering a Schwarzschild background of mass \(M_{\rm ADM} = H^\mathrm{real}\), the tail term \(T_{\ell m}\) is a resummed version of an infinite number of logarithmic terms that enter the transfer function between the nearÐ-zone and far-Ðzone waveforms. Since this complex object only resums the leading logarithms of tail effects, one needs to introduce an additional dephasing factor \(\delta_{\ell m}\), which is related to subleading logarithms.  The final building block is given by \(\left(\rho_{\ell m}\right)^\ell\), which was introduced to enhance the agreement of the EOB model with NR in the strong-field regime.  The explicit expressions for these various quantities can be found in Eqs. (19)-(21) and Appendix B of \cite{buho}.

 Having described the building blocks of the EOB model, we will now describe how to go about in the actual construction of the EOB model using NR simulations.  The first step in the calibration of the EOBNR model consists of aligning the waveforms at low frequency, following the procedure outlined in \cite{buho}. This approach is used to minimize the phase difference between the NR and EOB \((\ell,m)\) modes using the prescription

 \begin{equation}
 \varUpsilon (\Delta t, \Delta \phi)= \int^{t_2}_{t1} \left(\phi^{\rm EOB}(t+\Delta t) + \Delta \phi - \phi^{\rm NR}(t)\right)^2 \, dt,
 \label{alignment}
 \end{equation}
 
 \noindent where \(\Delta t/\Delta \phi\) are time/phase shifts, respectively, over which the minimization is performed. The time window \((t_1,t_2\)) is chosen so as to maximize the length of the NR waveform, but making sure that junk radiation does not contaminate the numerical data.  
 
The  numerical \(h_{22}\) is usually called the leading multipolar waveform because, compared to all the other multipoles, it provides the leading contribution to the amplitude of the full waveform \(h(t)\). Additionally, once the NR and EOB \((\ell,m)=(2,2)\) modes are aligned using the prescription given by Eq.~\eqref{alignment}, the peak of the numerical  \(h_{22}\) takes place at the same time the orbital frequency  \(\hat{\Omega}\) reaches its peak. Put in different words, the time at which the numerical \(h_{22}\) reaches its maximum and the EOB light-ring time (innermost circular orbit for a massless particle) are coincident. 

To calibrate the EOB dynamics, determined by Eqs.~\eqref{eq:eobhamone}-\eqref{eq:eobhamfour}, one minimizes the phase difference between the leading NR and EOB \((\ell,m)=(2,2)\) modes during the inspiral phase. This phase minimization procedure enables us to constrain the value of the doublet (\(a_5,\,a_6\)). As pointed out in \cite{buho}, the calibration of the adjustable parameters   (\(a_5,\,a_6\)) is not unique. For instance, \cite{buho} and \cite{rev} provide different values for the doublet \((a_5,a_6)\) which reproduce with great accuracy NR simulations for equal and comparable-mass non-spinning BH binaries. Even if the calibration of these parameters is degenerate, one can instill some physics in the way   (\(a_5,\,a_6\)) are determined.  In \cite{buho}, these parameters are modeled  as smooth functions of \(\eta\), such that they reproduce the self-force prediction of the orbital frequency shift at the ISCO in the test-mass particle limit \(\eta \rightarrow 0\), i.e., for  \(a_6(\eta\rightarrow 0)/\eta\) and  \(a_5(\eta\rightarrow 0)/\eta\) the EOB model reproduces the result \cite{inner}

\begin{equation}
M \Omega_{\rm ISCO} = \frac{1}{6\sqrt{6}}\left(1+1.2512\eta + {\cal{O}}(\eta^2)\right).
\label{ishi}
\end{equation}

It is worth pointing out that in contrast with the model developed in~\cite{tara}, the development of the EOBNRv2 model did not require the computation of higher-order PN corrections in the adjustable parameters \(\rho_{22}\) and \(\delta_{22}\). The actual expression for these two parameters had already been determined analytically in previous studies \cite{rev}. 

Having determined the EOB dynamics, the \(N_{22}\) coefficients are computed and implemented in the energy flux resummed prescription given by Eq.~\eqref{resflux}, one can determine the rest of the EOB adjustable parameters, namely, higher-order PN corrections in \(\rho_{\ell m}\)/\(\delta_{\ell m}\) for \((\ell,m) \neq (2,2)\), by minimizing the amplitude/phase difference between the remaining numerical and EOB multipolar waveforms used in the calibration. These higher-order corrections for \((\ell,m) \neq (2,2)\) are included in the inspiral waveform prescription, Eq.~\eqref{hip}, but not in the energy flux, Eq.~\eqref{resflux}.

\subsection{Merger and ring-down calibration of the EOB model}

The merger of two non spinning BHs generates a distorted Kerr BH whose gravitational radiation can be modeled using a superposition of quasi normal modes (QNMs), which are described by the indices (\(\ell,\, m,\,n\)), where \((\ell,\,m\)) denote the mode and \(n\) specifies the tone.  Each of these modes has a complex frequency \(\sigma_{\ell m n}\) given by

\begin{equation}
\sigma_{\ell mn} = \omega_{\ell m n} -i/\tau_{\ell m n},
\label{qnms}
\end{equation}

\noindent where the real/imaginary part \( \omega_{\ell m n}/\tau_{\ell m n}^{-1}\) corresponds to the frequency/inverse damping time of each QNM. These two observables are uniquely determined by the mass and spin of the Kerr BH formed after merger \cite{rdeq}. The prescription used to compute these quantities is given by  \cite{buho}

\begin{subequations}
\label{finalMS}
\begin{eqnarray}
    \!\!\!\!\!\!\frac{M_f}{M} &=& 1+\left (\sqrt{\frac{8}{9}}-1\right )\eta-0.4333 \eta^2-0.4392 \eta^3, \\
    \!\!\!\!\!\!\!\!\frac{a_f}{M_f} &=& \sqrt{12}\eta-3.871 \eta^2+4.028 \eta^3.
\end{eqnarray}
\end{subequations}

It is worth pointing out that a mode \((\ell,m\)) always consists of a superposition of two different frequencies/damping times. These `twin modes' are given by \( \omega'_{\ell m n} = -  \omega_{\ell -m n}\) and \(\tau'_{\ell m n} = \tau_{\ell -m n}\).  However, when considering two initially non-spinning BHs, the mirror solutions are degenerate in the modulus of the frequency and damping time, and hence one has that \( \omega_{\ell m n}>0\) and \( \tau_{\ell m n} >0\). 

Following \cite{buho}, the merger-ringdown waveform may be written as follows 

\begin{equation}
  \label{RD}
  h_{\ell m}^\mathrm{merger-RD}(t) = \sum_{n=0}^{N-1} A_{\ell mn}\,e^{-i\sigma_{\ell mn} (t-t_\mathrm{match}^{\ell m})},
\end{equation}
 
 \noindent where \(N\) is the number of overtones included in the model, i.e., \(N=8\), and \(A_{\ell mn}\) are complex amplitudes which will be determined by smoothly matching the inspiral-plunge waveform (Eq.~\eqref{hip}) with its merger ring-down counterpart (Eq.~\eqref{RD}).  
 
 To determine the complex amplitudes \(A_{\ell mn}\), one defines $t_\mathrm{match}^{\ell m}$ as the time at the amplitude maximum of the $h^{\rm EOB}_{\ell m}$ mode, namely, $t_\mathrm{match}^{\ell m}=t_\mathrm{max}^{\Omega}+\Delta t_\mathrm{max}^{\ell m}$, and demand continuity of the waveform at \(N-2\) points sampled in the time range \([ t_\mathrm{match}^{\ell m} - \Delta t_\mathrm{match}^{\ell m},\, t_\mathrm{match}^{\ell m}]\), and ensure the continuity and differentiability of the waveforms at \(t_\mathrm{match}^{\ell m} - \Delta t_\mathrm{match}^{\ell m}\) and \(t_\mathrm{match}^{\ell m}\)  (see Eqs. (36a)-(36c) in \cite{buho}).
 
Finally,  the full waveform can be written as

 \begin{equation}
  \label{eobfullwave}
  h_{\ell m} = h_{\ell m}^\mathrm{insp-plunge}\, {\mathcal H}(t_\mathrm{match}^{\ell m} - t) + h_{\ell m}^\mathrm{merger-RD}\,{\mathcal H} (t-t_\mathrm{match}^{\ell m})\,.
\end{equation}
 
 \noindent where \( {\mathcal H}(t)\) is the Heaviside step function. 
 
 This is the model we shall use in the following Section to develop an IMRI waveform model to explore the form of the self-force in the intermediate-mass ratio regime. We will consider three different systems with mass ratios 17:100, 10:100 and 1:100. We have chosen these systems because: a) the EOBNRv2 model was calibrated by comparison to NR simulations for events with mass-ratio 1:1-1:6, and hence systems with component masses 17:100 should be accurately modeled using EOBNRv2; b) the actual construction of the EOB model is such that it reproduces the expected dynamics of systems with small mass-ratio, e.g., \(\sim\) 1:100. This modeling statement will not be taken for granted in our subsequent analysis. We will show in the following Section that this is indeed the case by comparing results between EOBNRv2 and those obtained using Teukolsky data; c) we will explore an additional case, 1:10, for which the EOBNRv2 is the best model currently available to shed light on the form that self-force corrections should have so as to reproduce the dynamical evolution of these type of GW sources. Studying these type of sources is also important to try to bridge the gap in the parameter space covered by current waveform models.

 \section{Self-force corrections in the intermediate-mass-ratio regime}
 \label{s2}
 
 In the previous Section, we described the model we shall now use to build a waveform model for intermediate-mass-ratio inspirals (IMRIs). Using this IMRI model we will explore whether the inclusion of available self-force corrections, which have been computed in the extreme-mass-ratio regime~\cite{baracknewphi,sago,inner}, are able to reproduce the inspiral evolution of intermediate-mass-ratio binary black holes.

 \subsection{IMRI self-force model}
 
\begin{figure*}[ht]
\centerline{
\includegraphics[height=0.38\textwidth,  clip]{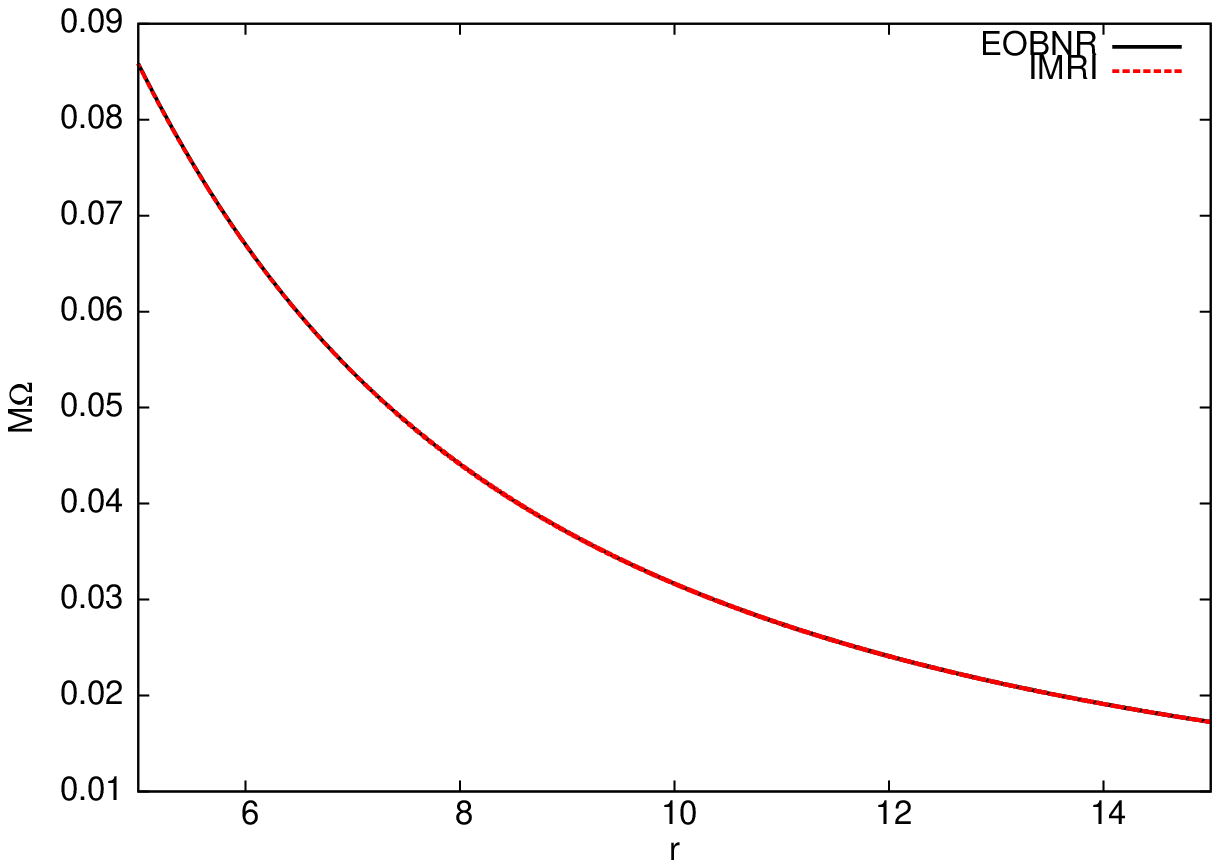}
\includegraphics[height=0.38\textwidth,  clip]{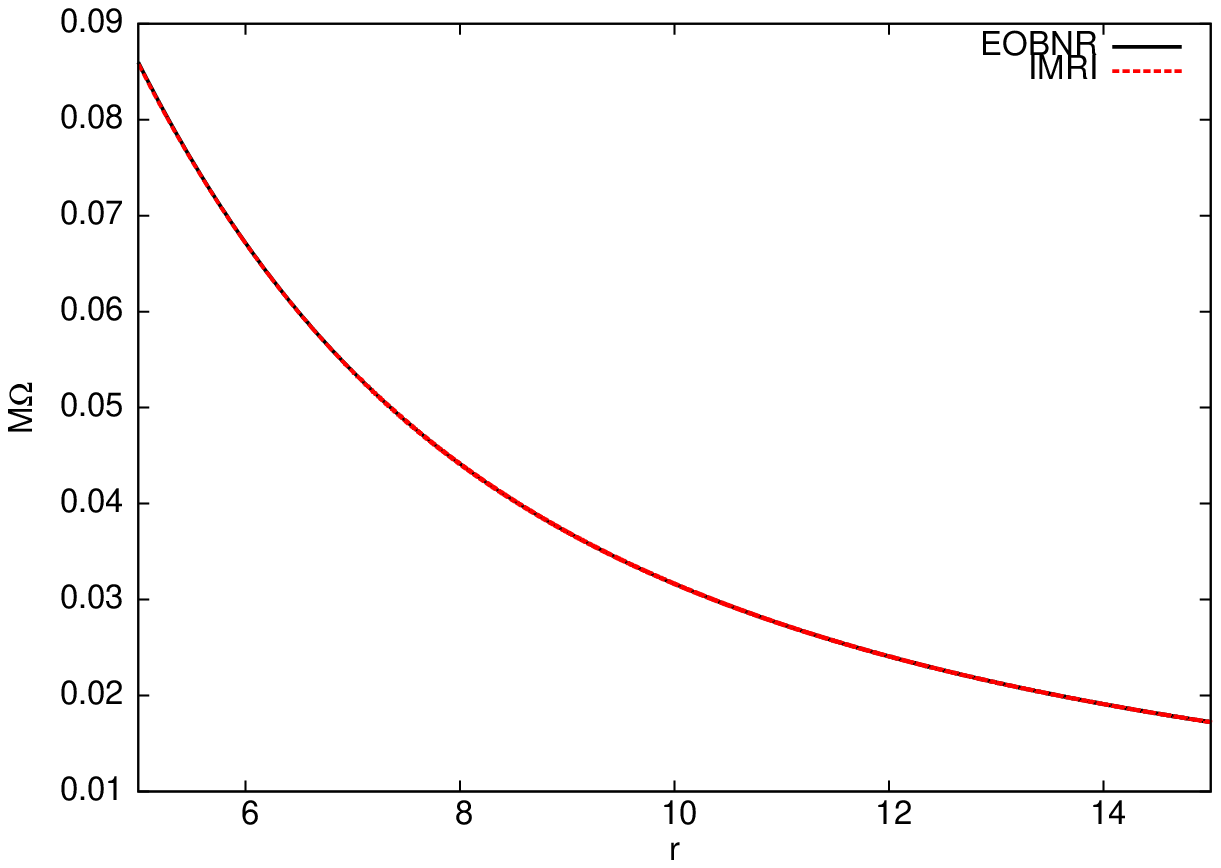}
}
\centerline{
\includegraphics[height=0.38\textwidth,  clip]{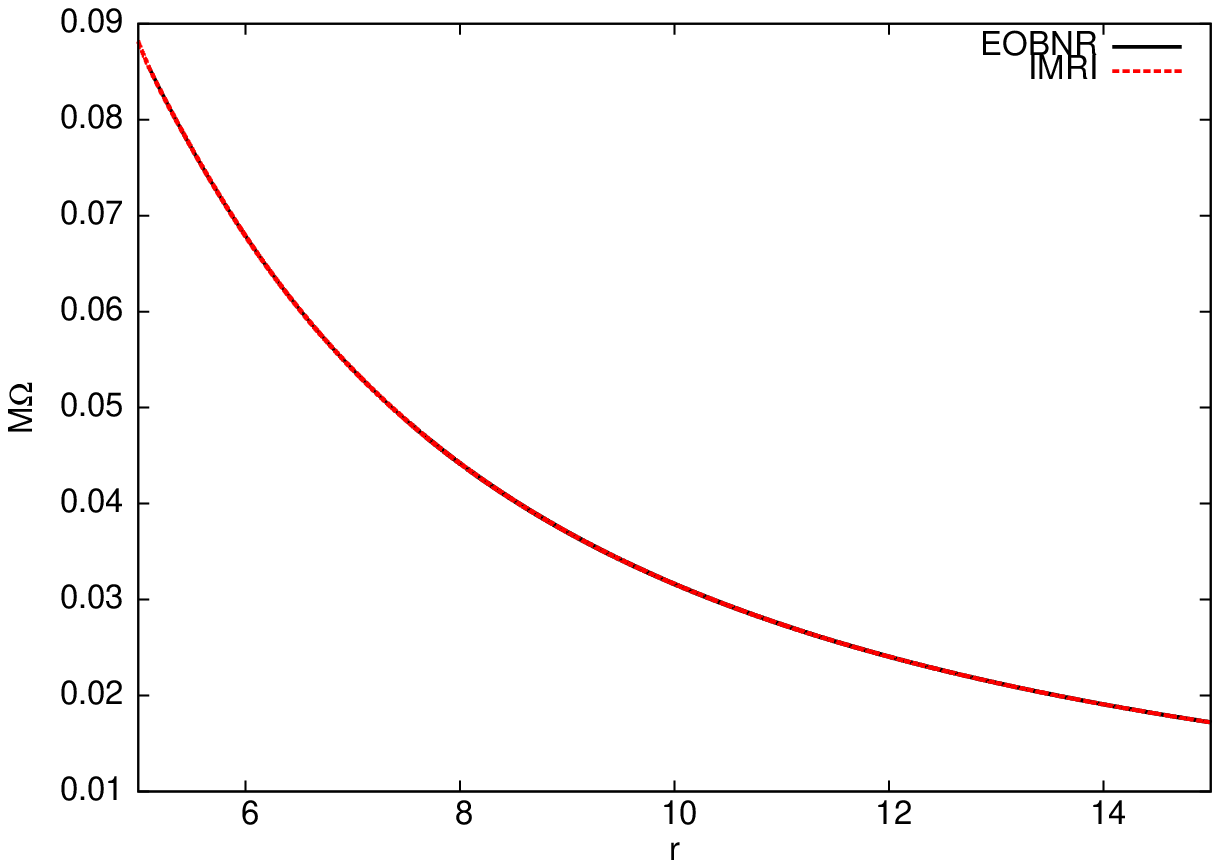}
\includegraphics[height=2.73in, width=3.75in,  clip]{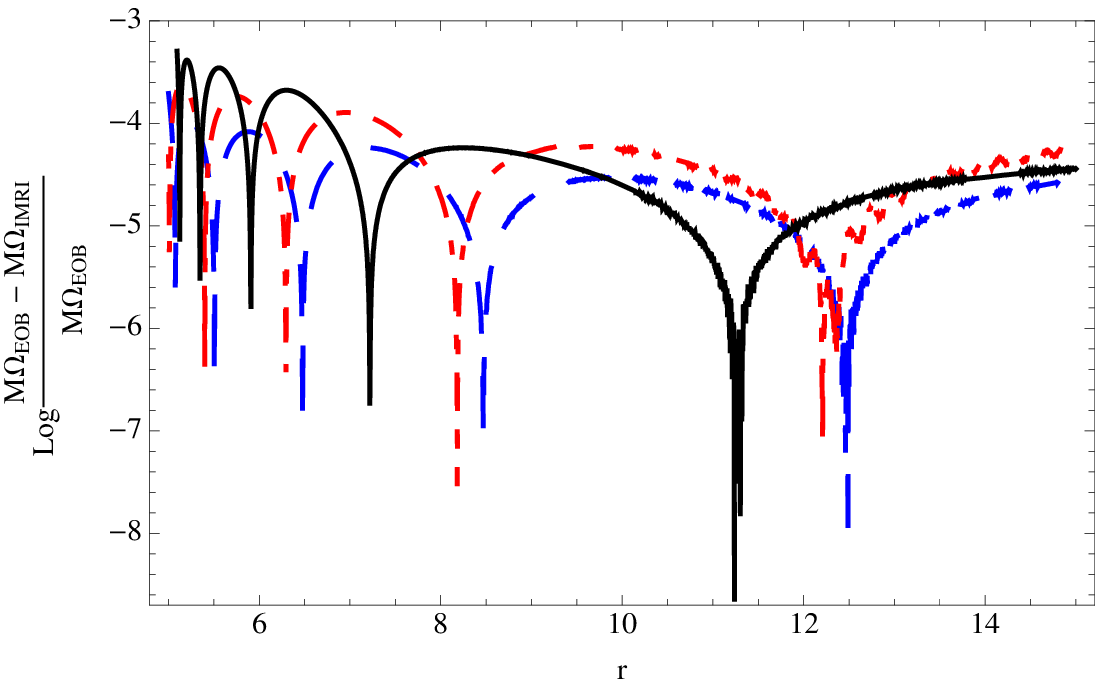}
}
\caption{The panels show the orbital frequency, as a function of the radial coordinate \(r\),  obtained using the fit described in the text (solid red line), and the orbital frequency predicted by the EOB model (solid black line). The systems shown in the panels correspond to binaries of component masses \(17 M_{\odot}+100M_{\odot}\) (top-left panel), \(10 M_{\odot}+100M_{\odot}\) (top-right) and  \(1 M_{\odot}+100M_{\odot}\) (bottom-left panel). The bottom-right panel demonstrates the accuracy with which our IMRI model reproduces the orbital frequency predicted by the EOBNRv2 for the systems 17:100 (dashed blue), 10:100 (dashed-dot red) and 1:100 (solid black). Note that the spikes are due to artifacts in the interpolation function used to plot the EOBNRv2 orbital frequency and the numerical fit used to reproduce it. Also notice that the discrepancy between the data and the fit is always smaller than one part in a thousand. }
\label{omegafit}
\end{figure*}

\begin{figure*}[ht]
\centerline{
\includegraphics[height=2.6in, width=3.8in,  clip]{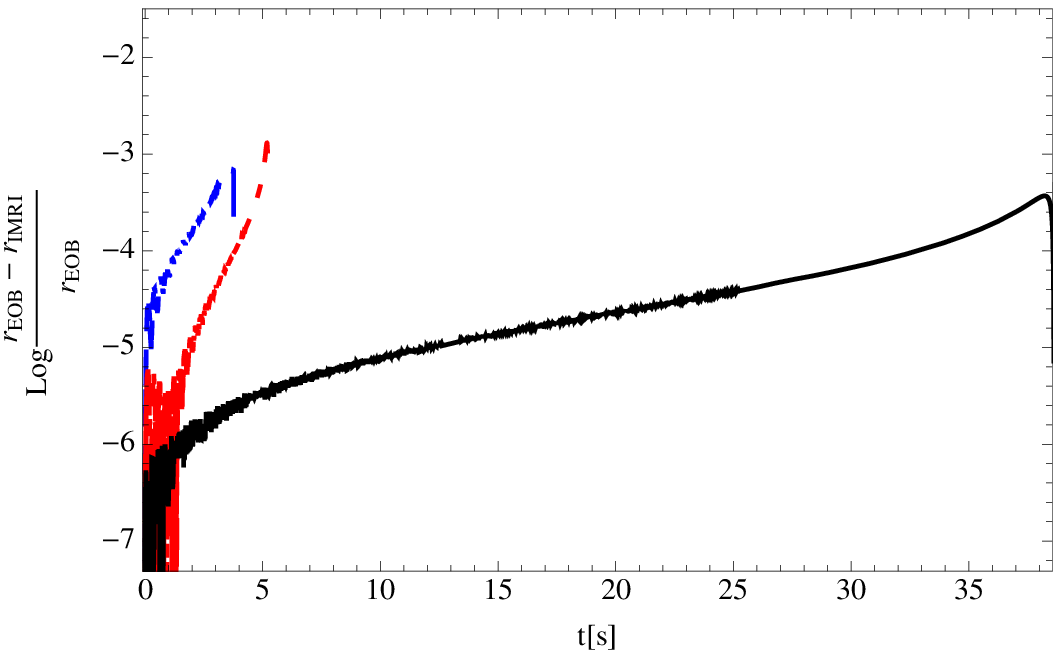}
\includegraphics[height=2.6in, width=3.8in,  clip]{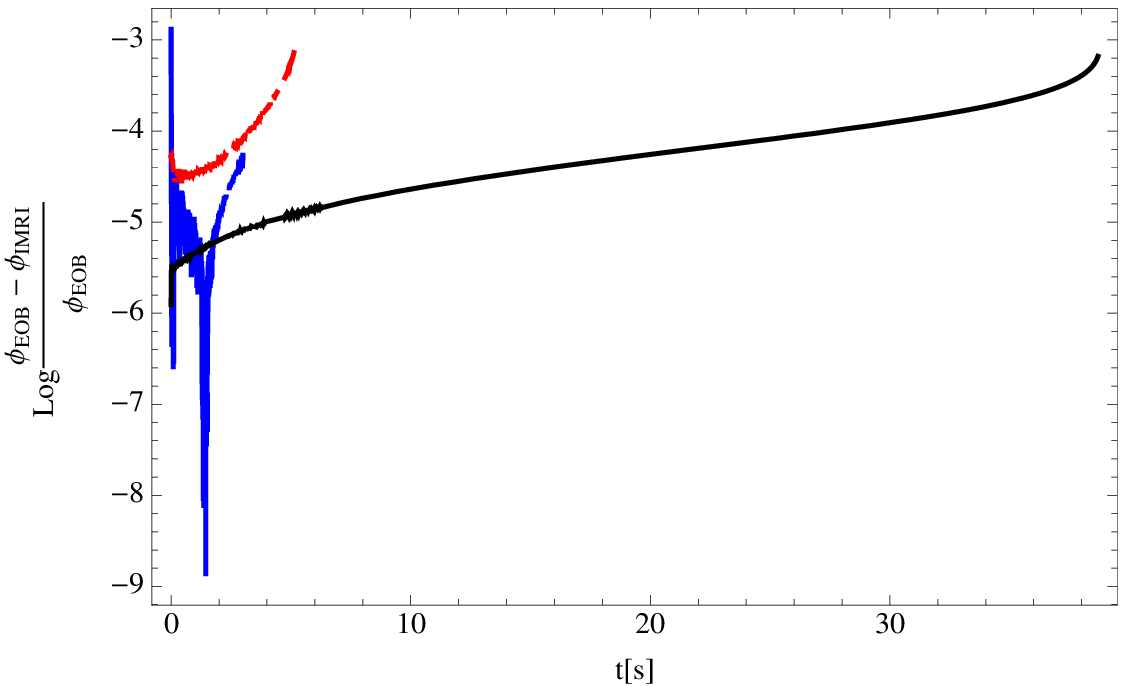}
}
\caption{The panels show the accuracy with which the IMRI model proposed in the paper reproduces the radial (right panel) and azimuthal (left-panel) time evolution predicted by the EOBNRv2 model for the systems  \(17M_{\odot} + 100M_{\odot}\) (dashed blue), \(10M_{\odot} + 100M_{\odot}\) (dashed-dot red), and \(1M_{\odot} + 100M_{\odot}\) (solid black). Notice that our model reproduces the EOBNRv2 orbital and azimuthal evolution point to point with an accuracy better than  one part in a thousand. }
\label{randphiaccuracy}
\end{figure*}

 In this Section we introduce the key elements we need to build our IMRI model which is simple and flexible enough to explore the form of the self-force in the intermediate-mass-ratio regime, and that, at the same time is able to reproduce with great accuracy the binary's dynamical evolution as predicted by EOBNRv2. In the following we assume that EOBNRv2 dynamics provides a good description of the actual binary black hole dynamical evolution. This is a reasonable assumption because, at present, EOBNRv2 is the best interface to translate NR simulations into a waveform model that uses coordinates which can be related to physical units~\cite{damsh}. This desirable modeling approach is obtained by comparing EOB gravitational waveforms to NR waveforms as seen by an observer at infinity~\cite{boyle}.  The fact that this model has been calibrated to NR simulations of mass-ratios 1:1-1:6 also means that such a model provides the adequate arena for the studies we want to carry out.  Having said that, we now describe the approach we shall follow to build our IMRI model:
 
 \begin{enumerate}
 \item We start with an ansatz for the orbital frequency evolution of our IMRI model which is inspired by recent studies on the inclusion of linear order self-force corrections in the EOB approach~\cite{barus}. We develop this prescription so as to faithfully reproduce the EOBNRv2 orbital frequency evolution (See Figure~\ref{omegafit}).  An accurate modeling of the orbital frequency is necessary to build the gauge-invariant object \(x=\hat\Omega^{2/3}\), which is a crucial element in our analysis. 
 \item We propose an ansatz to model the IMRI gravitational wave angular momentum flux which we calibrate so as to faithfully reproduce its EOBNRv2 counterpart (see Figures~\ref{fluxfitlight} and \ref{fluxfitheavy}). Note that the EOBNRv2 angular momentum flux used for this calibration  is constructed by summing over  35 leading and subleading waveform multipoles \(h_{\ell m}\), with  \(2\leq \ell \leq 8\), \(1\leq m\leq \ell\).  We include all these modes so as to model the radiative part of the self-force in our IMRI model as accurately as possible.
 \item Having derived an accurate prescription for the orbital frequency and the angular momentum flux, we make use of a prescription for the angular momentum that goes beyond the test-mass particle limit and which includes conservative self-force corrections. This expression for \(\hat{L}_z (x)\) encodes conservative self-force corrections in the redshift observable \(z_{\rm SF}\) (see Eq.~\eqref{zedsf}). 
 \item Because the prescriptions for the orbital frequency and the flux of angular momentum of our IMRI model reproduce their EOBNRv2 counterparts with great accuracy (see Figures~\ref{omegafit} and \ref{fluxfitlight}), we can now use these objects to constrain the form of the gauge-invariant expression of the angular momentum \(\hat{L}_z (x)\) in Eq.~\eqref{drdt} below by demanding internal consistency in our model, i.e., by reproducing faithfully the dynamical evolution of the binary black hole as predicted by EOBNRv2. This is equivalent to exploring the form of the red-shift observable \(z_{\rm SF}\) which reproduces the binary's dynamical evolution as predicted by EOBNRv2. This calibration procedure enable us to reproduce the expected inspiral evolution point to point to better than one part in a thousand, as shown in Figures \ref{dv15100} and \ref{randphiaccuracy}.
 \item Having determined the form of \(z_{\rm SF}\) that reproduces the dynamical evolution of the systems considered in this analysis, we compare the gauge-invariant expression of \(\hat{L}_z (x)\)  using our results and self-force corrections obtained in the context of EMRIs. We show that available self-force corrections do not reproduce the late inspiral evolution for systems with mass-ratio 1:6 and 1:10, but that they provide a fair description for systems with mass-ratio 1:100. This result is to be expected because these available self-force corrections were derived in the small mass-ratio limit.
 \end{enumerate}
 
 Having described the approach to be followed, we start off by describing the ansatz we use to model the orbital frequency. To do this,  we follow \cite{rev} and \cite{bernu}, and use the following prescription

\begin{equation}
\Omega=\frac{A}{\hat{H}}\frac{p_{\phi}}{r^2}.
\label{sOM}
\end{equation}

\noindent Furthermore, assuming circular orbits, we can simplify Eq.~\eqref{sOM} using the following relations for the conservative Hamiltonian $\hat{H}$

\begin{equation}
\hat{H}=\sqrt{A(u)\left(1+j_0^2 u^2\right)}, \quad {\rm with} \quad p_\phi=j_0\,, \quad {\rm and} \quad j_0^2=-\frac{A'(u)}{\left(u^2A(u)\right)'}\,,
\label{assum}
\end{equation}
 
\noindent where \(u=1/r\) and \('\) denotes \(d/du\). Using a prescription similar to that introduced in \cite{barus}, we shall use the following ansatz for the potential \(A(u)\)

\begin{equation}
A_{\rm ansatz}(u)= 1-2u+\eta\left( \sqrt{1-3u}\,k_{\rm fit} -u \left(1+\frac{1-4u}{\sqrt{1-3u}}\right)\right), \quad {\rm with} \quad k_{\rm fit} = 2u\frac{1+a_1u+a_2 u^2}{1+a_3 u +a_4 u^2 + a_5 u^3}\,.
\label{ansatz}
\end{equation}

\noindent Under these assumptions, the prescription for the orbital frequency takes the simple form

\begin{equation}
\Omega(u)_{\rm ansatz}= u^{3/2}\sqrt{-\frac{A'_{\rm ansatz}(u)}{2}}\,.
\label{ompres}
\end{equation}

\noindent The values of the \(a_i\) parameters in the function \(k_{\rm fit}\) are given in Table~\ref{coef}. It is worth pointing out that this prescription captures accurately the orbital phase evolution of the sources described above all the way down to ISCO and slightly beyond with much less computational complexity than EOBNRv2.  The actual comparison between this scheme and the EONBRv2 model is shown in Figure~\ref{omegafit}. Notice that our IMRI model does pretty well from large \(r\) to the fast-motion strong-field regime in all three cases shown. Notice that we are modeling the potential  \(A_{\rm ansatz}(u)\) using the coordinate \(u\) and not \(M/x\) as done in \cite{barus}. The rationale for doing this is to use a simple, yet accurate, prescription for the orbital frequency that captures faithfully the EOBNRv2 orbital frequency evolution. We could have also used a different prescription for this object, i.e., a PN series~\cite{cons}, a Pad\'e resummed expression, etc. The purpose this object serves at this stage in the analysis is to reproduce accurately the orbital evolution predicted by EOBNRv2 with less computational complexity.

\begin{table}[ht!]
\centering
\begin{tabular}{c c c c c c}
\hline\hline
  $\eta$& $a_1$ & $a_2$ & $a_3$ & $a_4$ & $ a_5 $  \\  
    $\frac{1700}{13689}$& -7.458 & 15.0179 & -7.208 & 12.152 & 6.103 \\ [1ex]
   $\frac{10}{121}$& -8.037 & 17.059 & -7.826 & 14.589 & 5.393\\ [1ex]
    $\frac{100}{10201}$&-9.535 & 22.984 & -9.456 & 21.862 & 3.085 \\ [1ex]
\hline
\end{tabular}
\caption{\(\Omega_{\rm ansatz}\) fit coefficients}
\label{coef}
\end{table}

Having derived a prescription for the orbital evolution, we require a prescription to generate the inspiralling trajectory of the stellar mass compact object. The first step to achieve this consists of deriving a consistent model for the flux of angular momentum. The model for the flux of angular momentum that we introduce in the following Section sums up the effect of including all the dominant and subdominant modes currently available in the literature, i.e., \(2\leq \ell \leq 8, \, 1\leq m \leq \ell\), with the advantage that the inclusion of extra modes does not impact the computational cost of generating the IMRI waveform.

\subsection{Radial evolution prescription}

An important component of the IMRI waveform model is the prescription used for the fluxes of energy and angular momentum. Deriving an accurate expression for these quantities is essential to capture the main features of true inspirals, as shown in \cite{improved} in the context of EMRIs.

As discussed in Section~\ref{s1}, in the EOBNRv2 model the flux of energy is constructed using the prescription given by Eq.~\eqref{resflux}, which is obtained by summing over the modes  \(( h_{\ell m}\)) with \(2\leq \ell \leq 8\), \(1\leq m\leq \ell\). Using this prescription as input data and the fact that in the EOB formalism the following relation is fulfilled~\cite{barus}

\begin{equation}
\dot E= \Omega\dot L_z,
\label{circr}
\end{equation}

\noindent we derive a prescription for the gravitational-wave angular momentum flux which is valid from early inspiral all the way to the ISCO and slightly beyond.  This prescription, which encapsulates the contribution from 35 dominant and subdominant modes, is inspired by the modeling of accurate EMRI waveform models introduced in~\cite{improved}, i.e., 

\begin{equation}
\left(\dot L_z\right)_{\rm fit} = -\frac{32}{5}\frac{\mu^2}{M}\frac{1}{r^{7/2}}\Bigg[1-\frac{1247}{336}\frac{1}{r}+4\pi\frac{1}{r^{3/2}}- \frac{44711}{9072}\frac{1}{r^2} + \frac{1}{r^{5/2}}\left(c^1_{2.5} +  c^2_{3}\frac{1}{r^{1/2}}+c^3_{3.5}\frac{1}{r}\right)\Bigg],
\label{lzfluxfit}
\end{equation}

\noindent and the coefficients in Eq.~\eqref{lzfluxfit} are given in Table~\ref{Lzdotcoef}.

\begin{table}[ht]
\centering
\begin{tabular}{c c c c}
\hline\hline
  $\eta$& $c^1_{2.5}$ & $c^2_{3}$ & $c^3_{3.5}$   \\  
    $\frac{1700}{13689}$& -121.903 & 551.141 &-694.699  \\ [1ex]
   $\frac{10}{121}$& -110.657 & 360.261 &-156.016 \\ [1ex]
    $\frac{100}{10201}$&-75.255 & 333.449 & -363.505  \\ [1ex]
\hline
\end{tabular}
\caption{\(\dot{L}_z\) fit coefficients}
\label{Lzdotcoef}
\end{table}

\begin{figure*}[ht]
\centerline{
\includegraphics[height=0.38\textwidth,  clip]{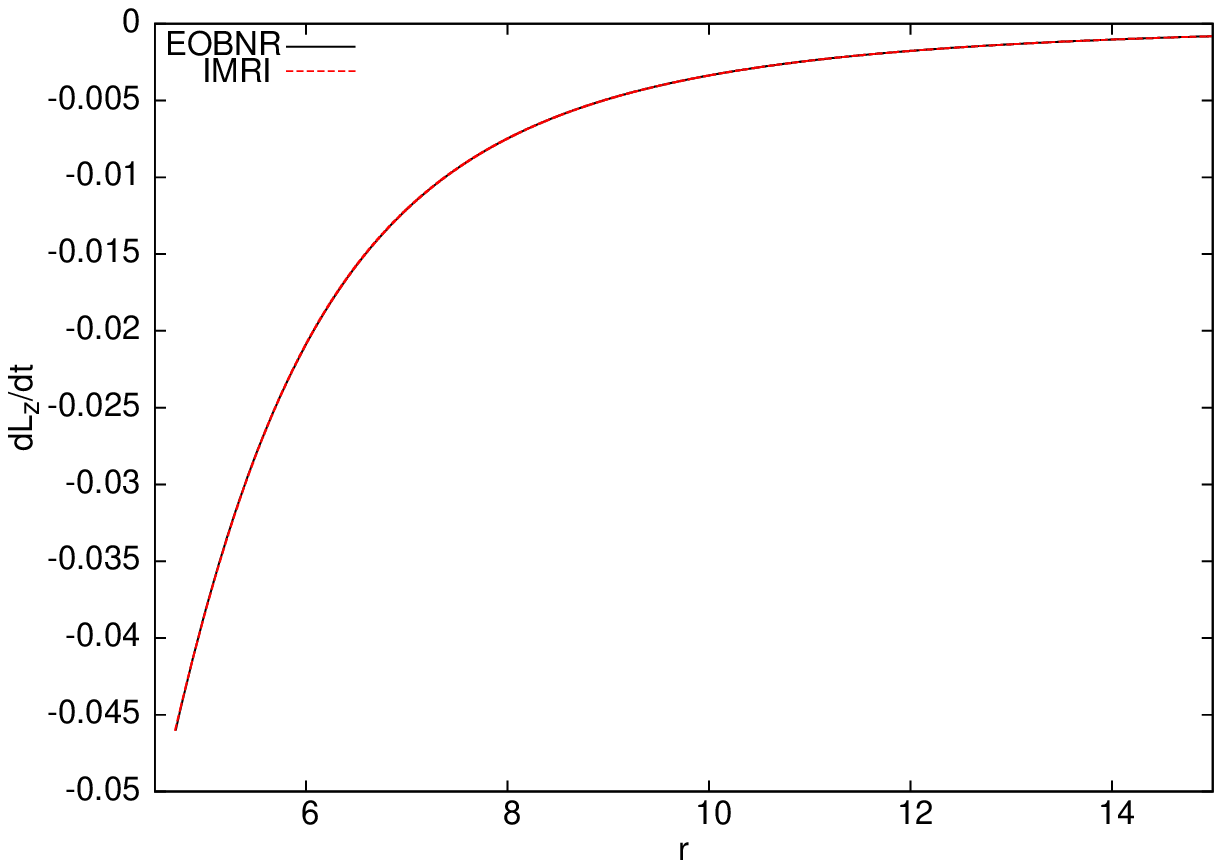}
\includegraphics[height=0.38\textwidth,  clip]{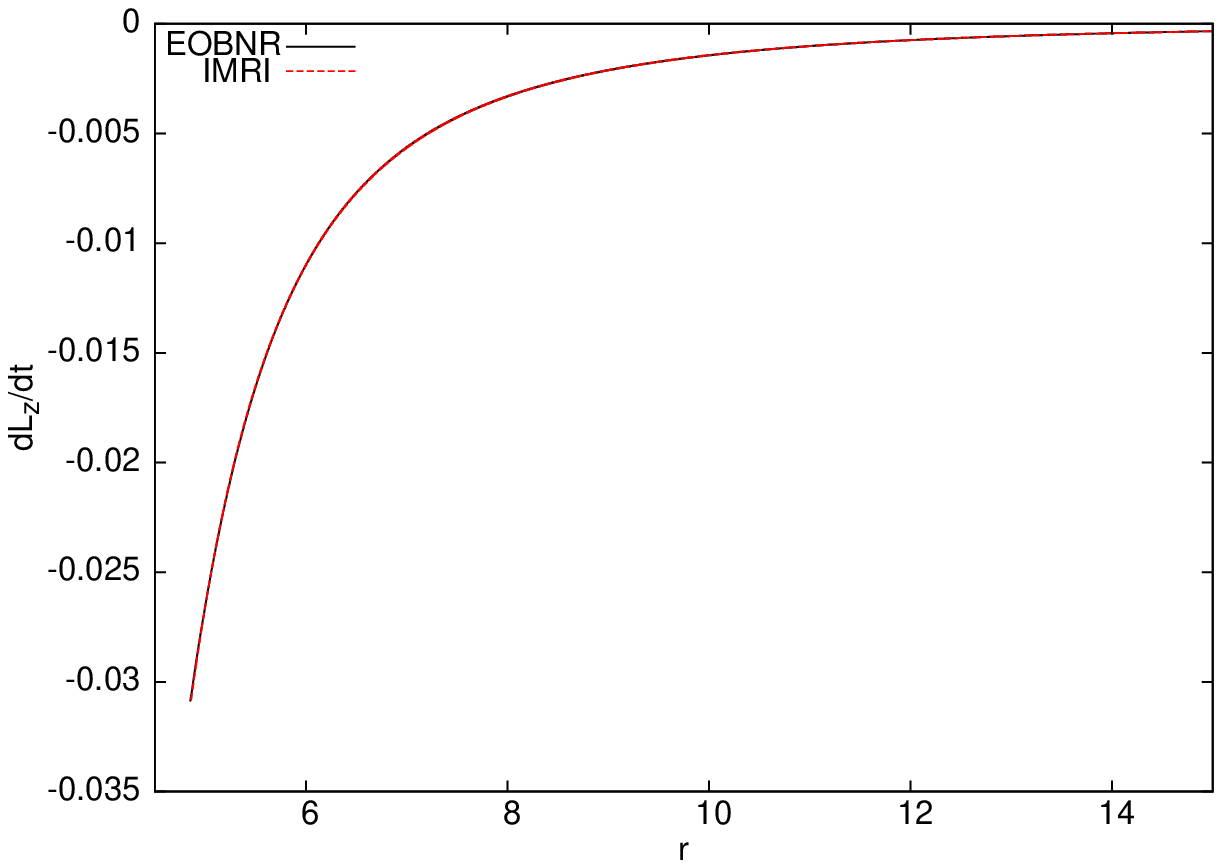}
}
\centerline{
\includegraphics[height=0.38\textwidth,  clip]{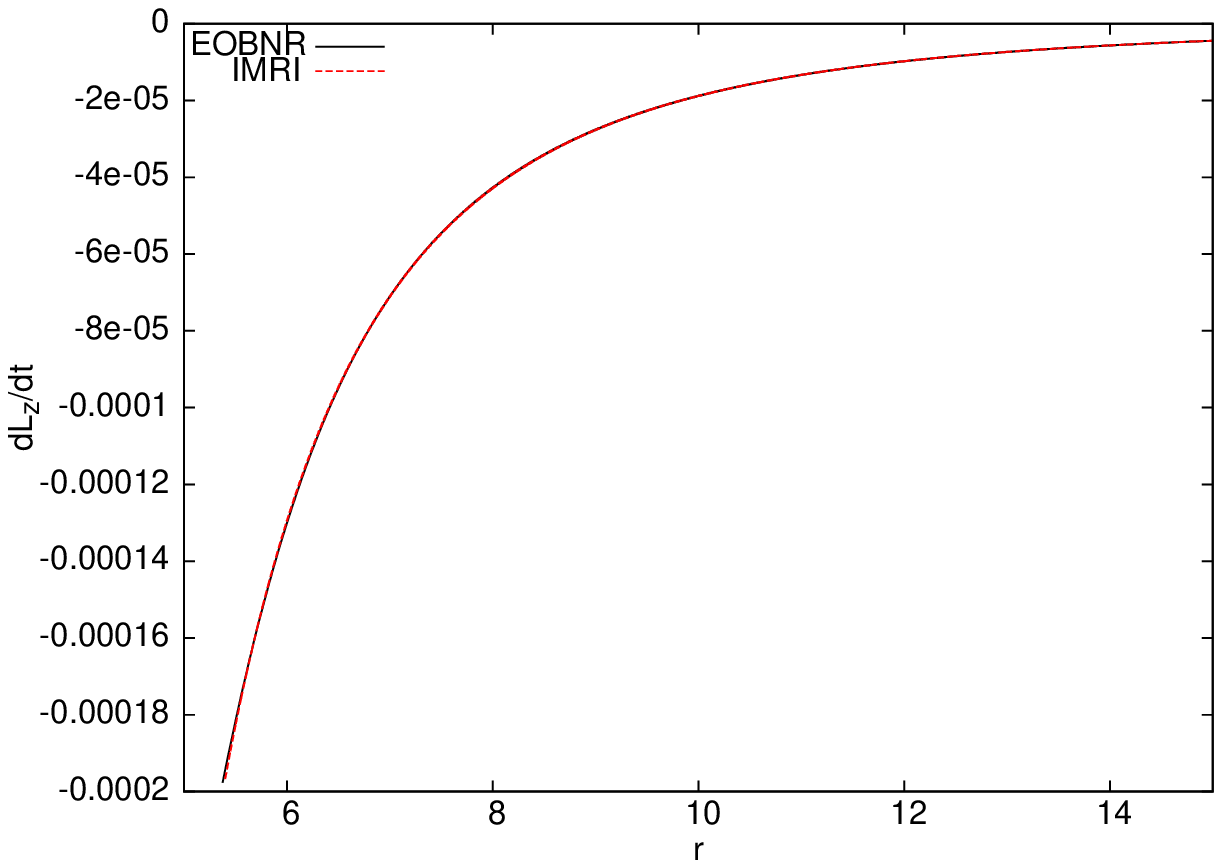}
\includegraphics[height=2.6in, width=3.8in,  clip]{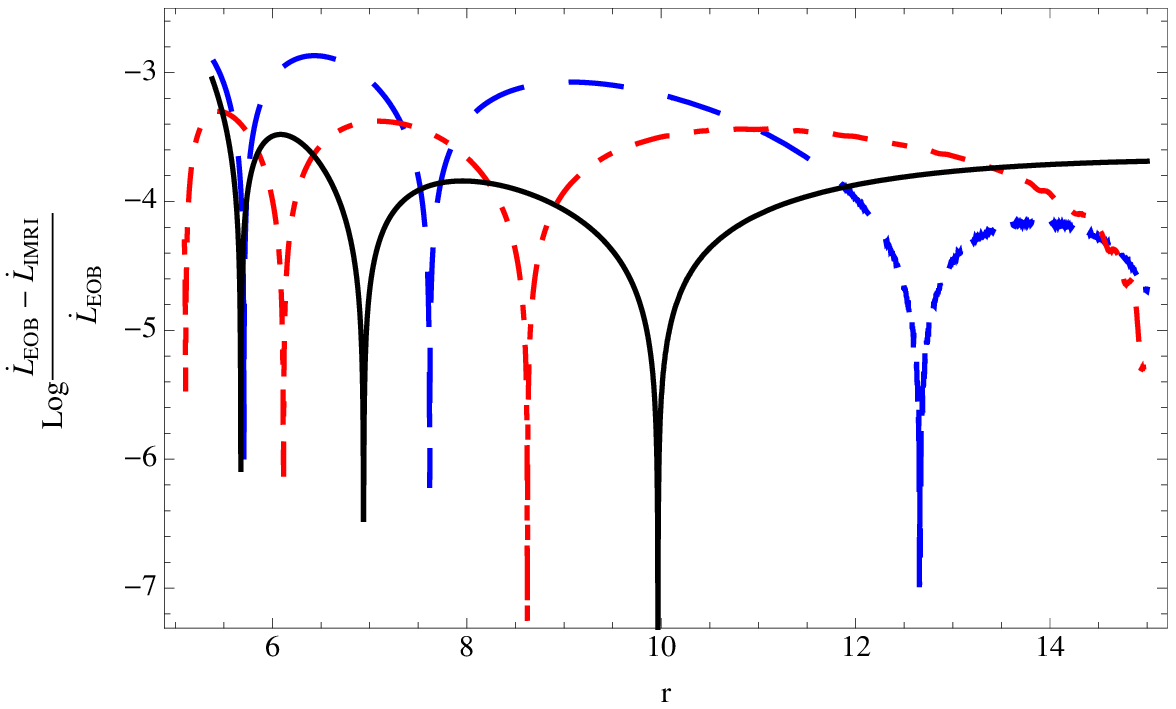}
}
\caption{We compare the calibrated flux of angular momentum described in Eq.~\eqref{lzfluxfit} against the prediction of the EOBNRv2 model for binaries of component masses \(17M_{\odot} + 100M_{\odot}\) (top left panel),  \(10M_{\odot} + 100M_{\odot}\) (top right panel), and \(1M_{\odot} + 100M_{\odot}\) (bottom-left panel). The bottom-right panel shows the accuracy with which the IMRI model reproduces the EOBNRv2 angular momentum flux, \(\dot{L}=dL_z/dt\), for the systems: 17:100 (dashed blue), 10:100 (dashed-dot red) and 1:100 (solid black). The spikes in the bottom-right panel are due to numerical artifacts of the interpolating function used to plot the EOBNRv2 angular momentum flux and the numerical fit to reproduce it. The fit is such that its discrepancy to the EOBNRv2  is always smaller than one part in a thousand.}
\label{fluxfitlight}
\end{figure*}

\begin{figure*}[ht]
\centerline{
\includegraphics[height=0.38\textwidth,  clip]{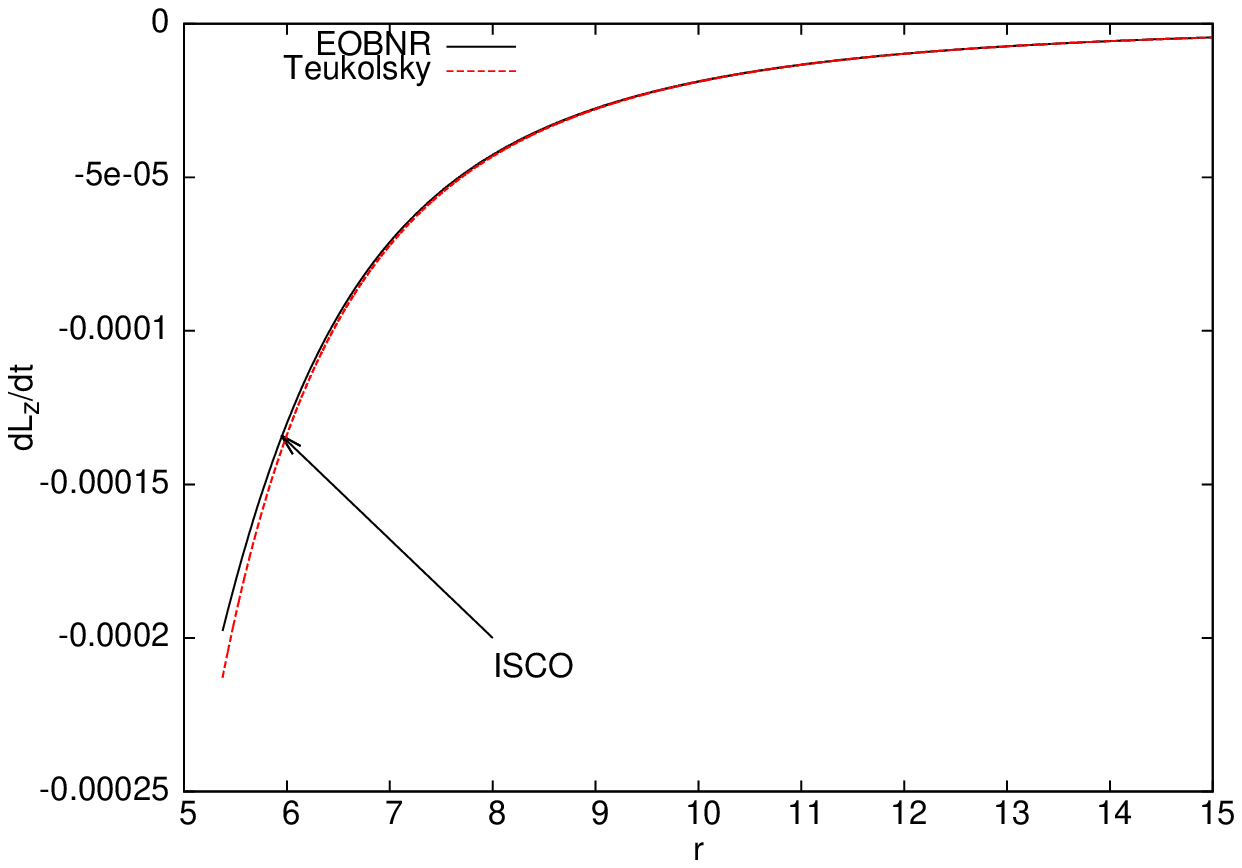}
\includegraphics[height=2.6in, width=3.8in,  clip]{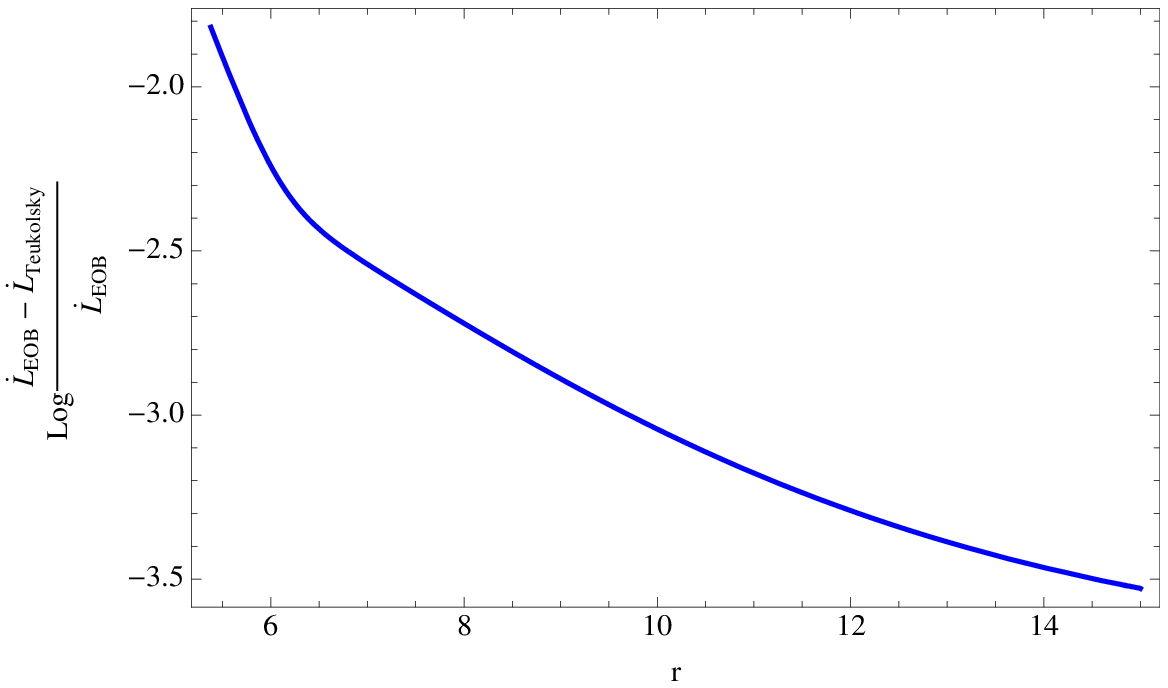}
}
\caption{The left panel shows the angular momentum flux for binaries of mass ratio 1:100 computed using the EOBNRv2 model and the fit to Teukolsky data introduced in \cite{improved}. Both formalisms present a remarkable agreement all the way down to the ISCO. Note that the angular momentum flux fit to Teukolsky data is valid only from early inspiral until the ISCO. The right panel shows the relative difference between the two prescriptions for the angular momentum flux \(\dot{L}=dL_z/dt\).}
\label{fluxfitheavy}
\end{figure*}

Figures~\ref{fluxfitlight} and \ref{fluxfitheavy} present the comparison between the EOBNRv2 flux, which includes all currently known dominant and subdominant modes, and the calibrated angular momentum flux \(\left(\dot L_z\right)\). Notice that this computationally inexpensive scheme does pretty well from large \(r\) all the way to the ISCO and slightly beyond. Hence, this approach will enable us to capture the main features of the inspiral evolution of the systems under consideration in the regime of interest with good accuracy. Modelling the prescription of the flux of angular momentum in our IMRI waveform using all the dominant and subdominant modes in the EOBNRv2 model is equivalent to modeling the radiative part of the self-force with the best information currently available in the literature.

In Figure~\ref{fluxfitheavy} we present a comparison between the flux of angular momentum predicted by the EOBNRv2 model and the fit to Teukolsky data proposed in \cite{improved} for extreme-mass-ratio inspirals. Note the remarkable agreement between both formalisms all the way down to the ISCO. This comparison also confirms that the construction of the EOBNRv2 captures the main features of inspirals with small mass-ratios which are modeled using black hole perturbation theory. This comparison also shows that the EOBNRv2 model encodes fairly well the radiative part of the self-force for systems with small mass-ratio.

Having found a prescription for the angular momentum that captures the dynamics of IMRIs, we can now use it to generate the inspiral trajectory of the stellar mass compact object that inspirals into an IMBH using the relation 

\begin{equation}
\frac{d r}{d t}= \frac{d L_z}{d t}\frac{d r}{d L_z}.
\label{drdt}
\end{equation}

\noindent  The first term on the right hand side of Eq.~\eqref{drdt} can be obtained from Eq.~\eqref{lzfluxfit}. With regard to the derivative of the angular momentum with respect to the radial coordinate, we need to use a prescription for the angular momentum that goes beyond the test-mass particle limit, as conservative self-force corrections play a more significant role in this regime. Hence, we use the relation derived by Barausse et al in \cite{barus} which includes conservative self-force corrections, i.e.,

\begin{equation}
\hat{L}_z (x)= \frac{L_z}{\mu M}= \frac{1}{\sqrt{x(1-3x)}} + \eta\left(-\frac{1}{3 \sqrt{x}} z'_{\rm SF}(x) + \frac{1}{6\sqrt{x}}\frac{4-15x}{\left(1-3x\right)^{3/2}}\right) + {\cal O} (\eta^2),
\label{angmom}
\end{equation}

\noindent where \(x=(M\Omega)^{2/3}\) and \('\) stands for \(d/dx\). Furthermore,

\begin{equation}
z_{\rm SF} (x)= 2x \frac{1+b_1 x+b_2 x^2}{1+b_3x +b_4 x^2 + b_5 x^3},
\label{zedsf}
\end{equation}

\noindent and the various coefficients \(b_i\) were derived in \cite{barus} using available self-force data. In this section we derive the coefficients in Eq.~\eqref{zedsf} that reproduce the actual inspiral evolution all the way down to the ISCO. Note that in Eq.~\eqref{drdt} we have in place a prescription for the angular momentum flux which faithfully reproduces the expected loss of angular momentum even beyond the ISCO,  as compared to the EOBNRv2 model. Hence, the only ingredient that needs to be tuned to reproduce the expected inspiral evolution is contained in Eq.~\eqref{angmom}. We have followed this approach to explore the form that \(\hat{L}_z (x)\) should have. Put in different words, the form that the self-force redshift observable \(z_{\rm SF} (x)\) should have to reproduce an inspiral trajectory consistent with EOBNRv2 all the way down to the ISCO.  The various coefficients of Eq.~\eqref{zedsf} that generate such an inspiral orbit are given in Table~\ref{zedfunccoef}.

\begin{table}[ht]
\centering
\begin{tabular}{c c c c c c}
\hline\hline
  $\eta$& $b_1$ & $b_2$ & $b_3$ & $b_4$ & $ b_5 $  \\  
    $\frac{1700}{13689}$& -3.062 & 0.760 & -3.261 & 0.550 & -6.000 \\ [1ex]
   $\frac{10}{121}$& -3.260 & 0.892 & -3.475 & 0.600 & -6.829\\ [1ex]
    $\frac{100}{10201}$&-3.370 & 1.570 & -3.730 & 0.970 & -7.280 \\ [1ex]
\hline
\end{tabular}
\caption{\(z_{\rm SF} (x)\) fit coefficients}
\label{zedfunccoef}
\end{table}

\begin{figure*}[ht]
\centerline{
\includegraphics[height=0.38\textwidth,  clip]{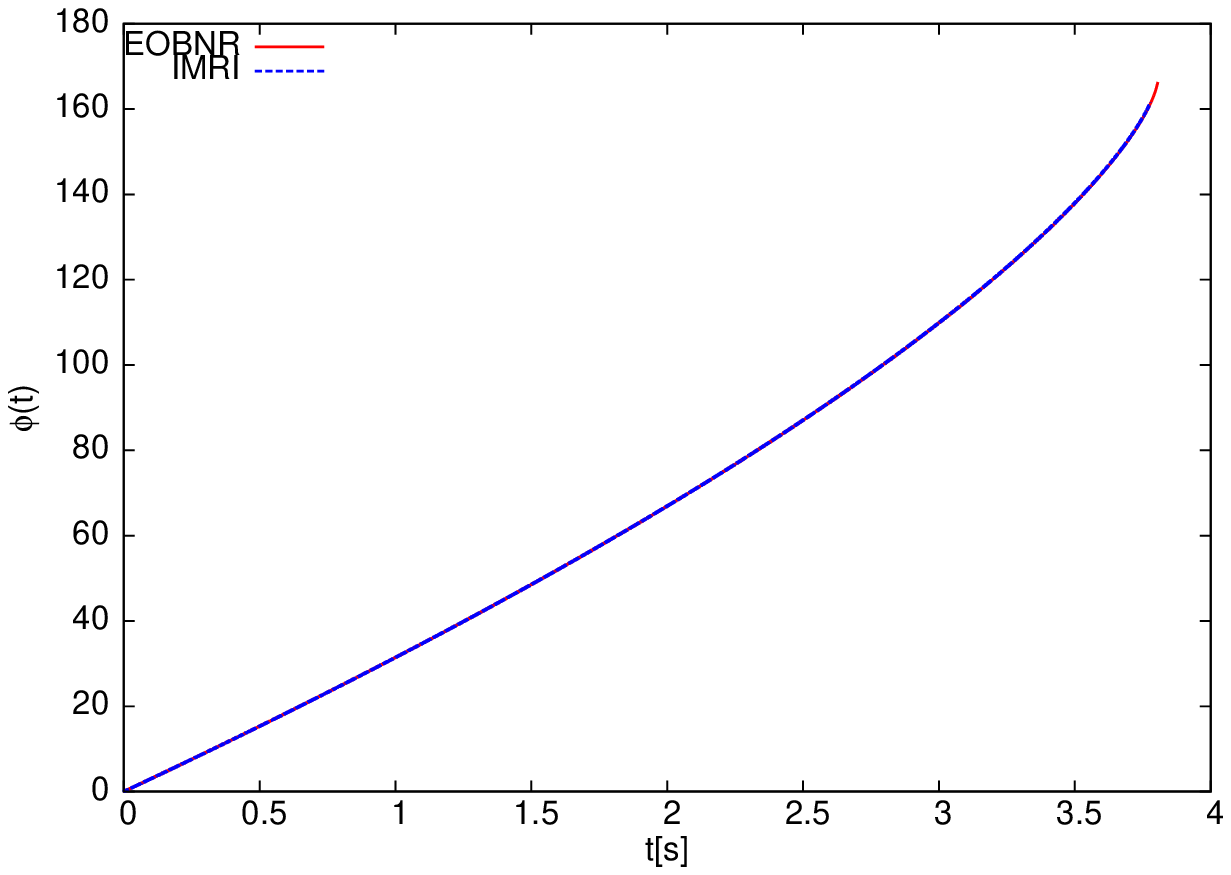}
\includegraphics[height=0.38\textwidth,  clip]{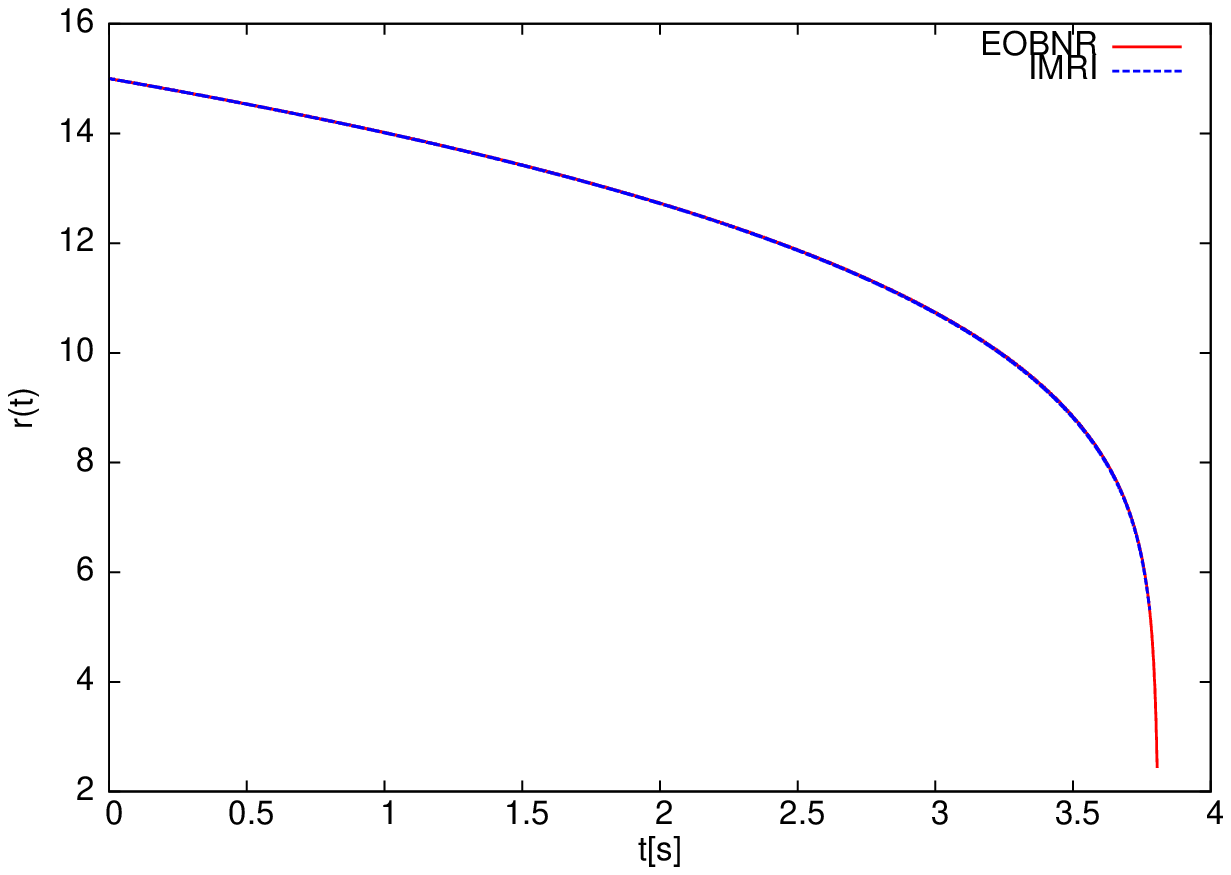}
}
\centerline{
\includegraphics[height=0.38\textwidth,  clip]{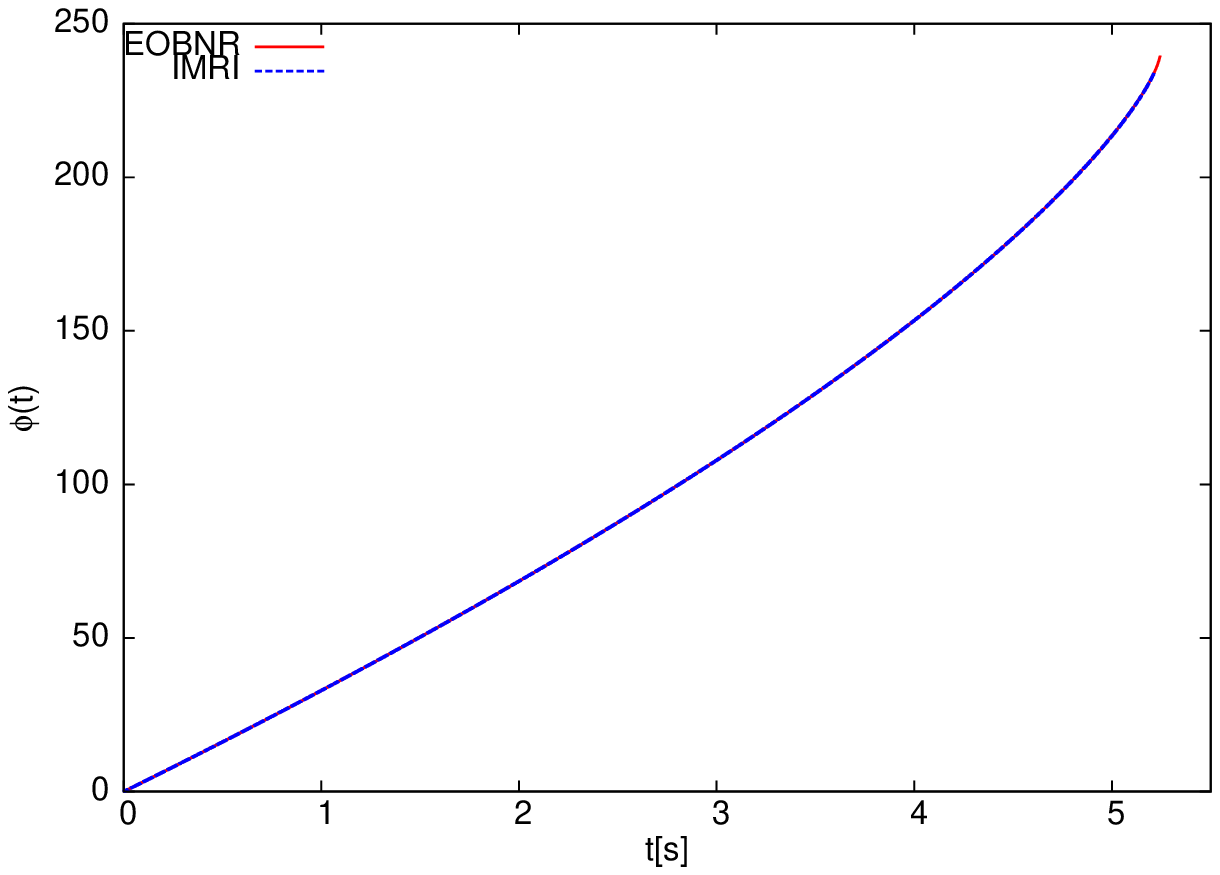}
\includegraphics[height=0.38\textwidth,  clip]{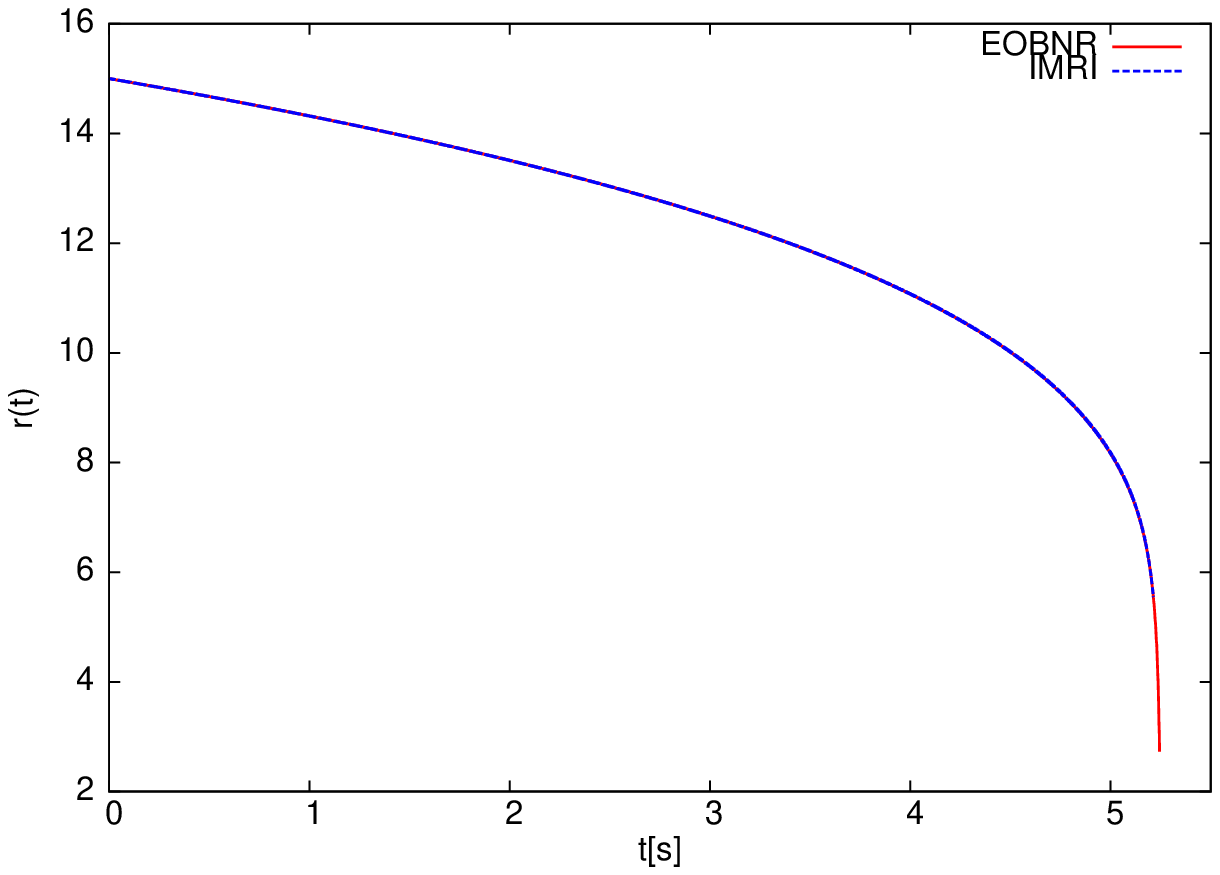}
}
\centerline{
\includegraphics[height=0.38\textwidth,  clip]{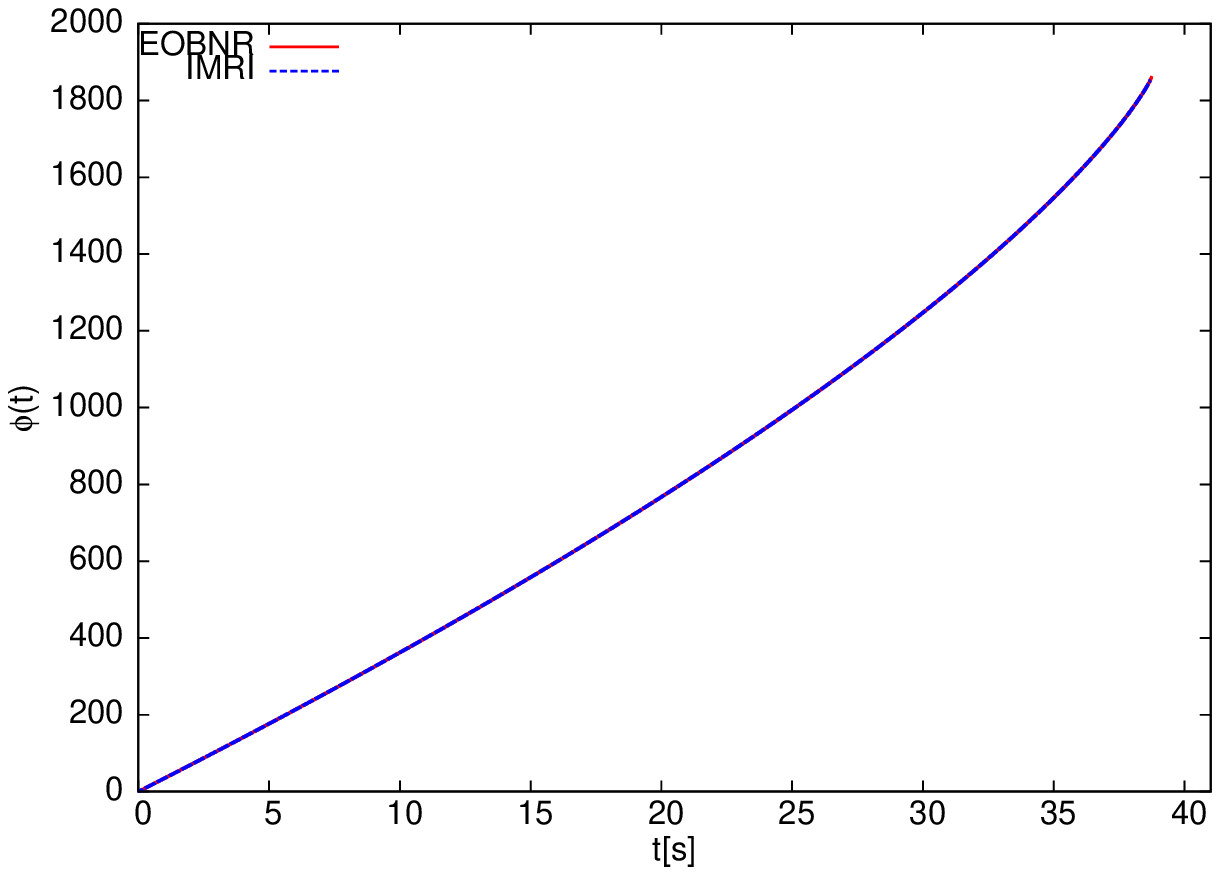}
\includegraphics[height=0.38\textwidth,  clip]{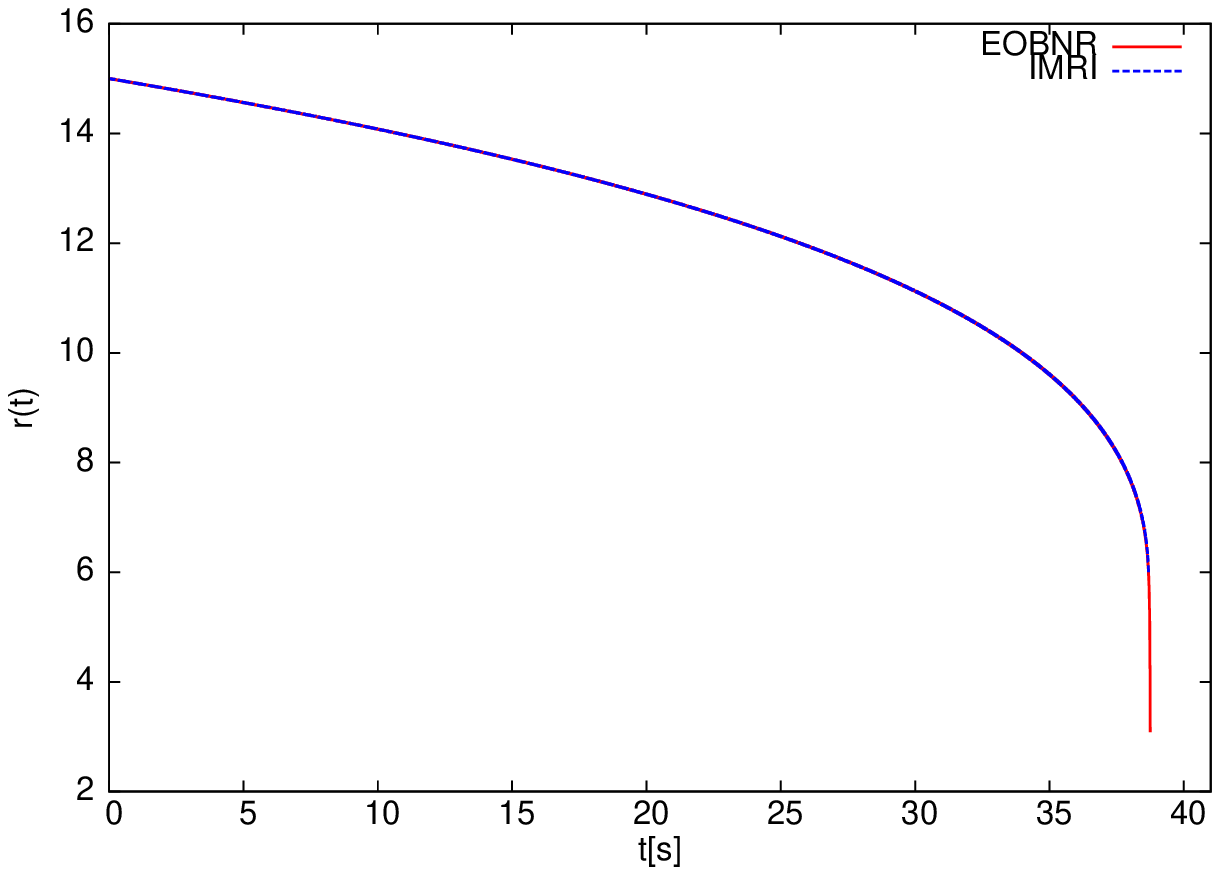}
}
\caption{The panels show the radial and azimuthal evolution for a \(17M_{\odot} + 100M_{\odot}\) system (top panels), \(10M_{\odot} + 100M_{\odot}\) system (middle panels), and \(1M_{\odot} + 100M_{\odot}\) system (bottom panels), obtained using the IMRI model compared against the EOBNRv2 model.}
\label{dv15100}
\end{figure*}

In Figure~\ref{dv15100} we show that this prescription captures with great accuracy the features of EOBNRv2  inspirals. To recap,  the IMRI model incorporates both first-order conservative corrections --through  the construction of the gauge invariant expression of the angular momentum in  Eq.~\eqref{angmom}--- and first-order radiative-corrections through the construction of the flux of angular momentum in Eq.~\eqref{lzfluxfit}. Notice that the latter object encodes the best information currently available in the literature, since it includes the contribution from 35 dominant and subdominant \((\ell,m)\) modes of the multipolar waveform to the angular momentum flux. Using this IMRI model, we have been able to constrain the form of the angular momentum \(L_z(x)\) that reproduces the inspiral evolution predicted by the EOBNRv2 model.

\begin{figure*}[ht]
\centerline{
\includegraphics[height=0.38\textwidth,  clip]{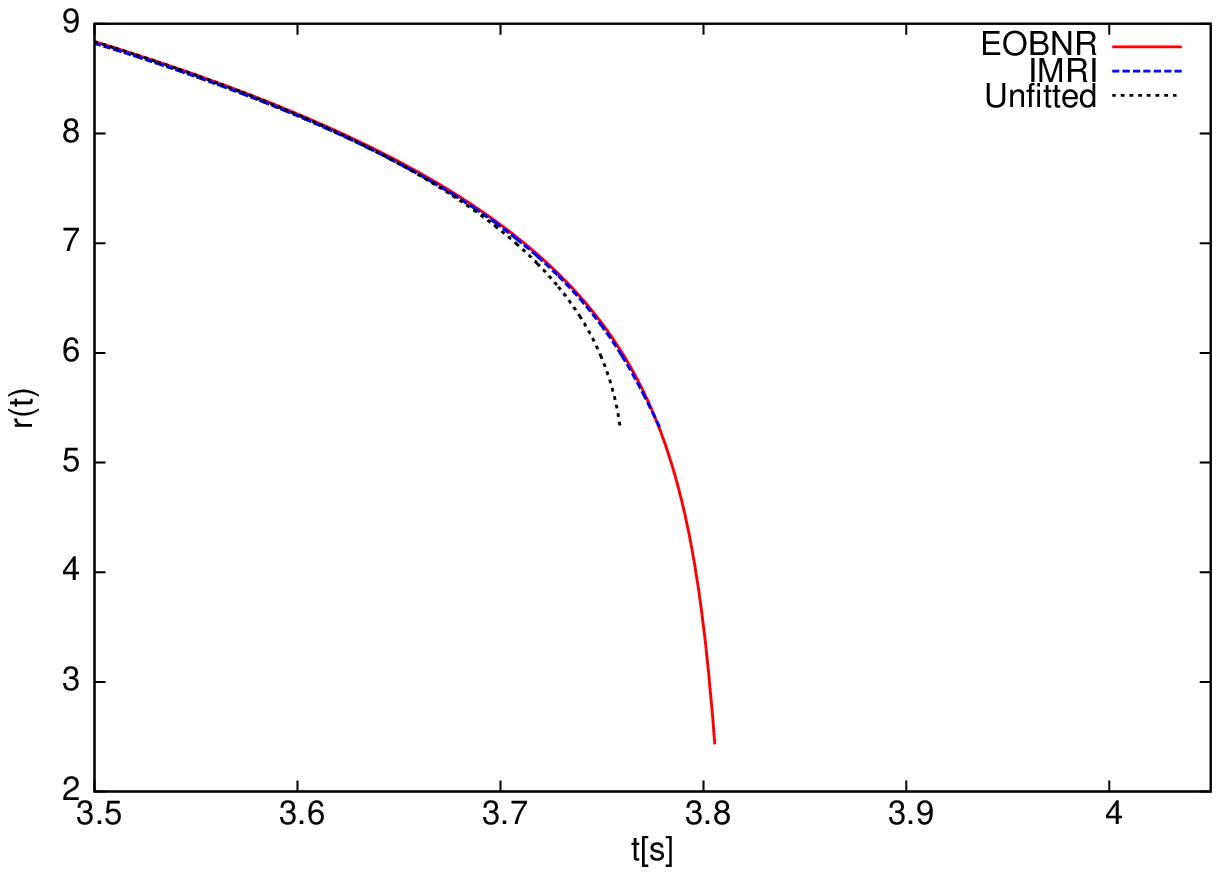}
\includegraphics[height=0.38\textwidth,  clip]{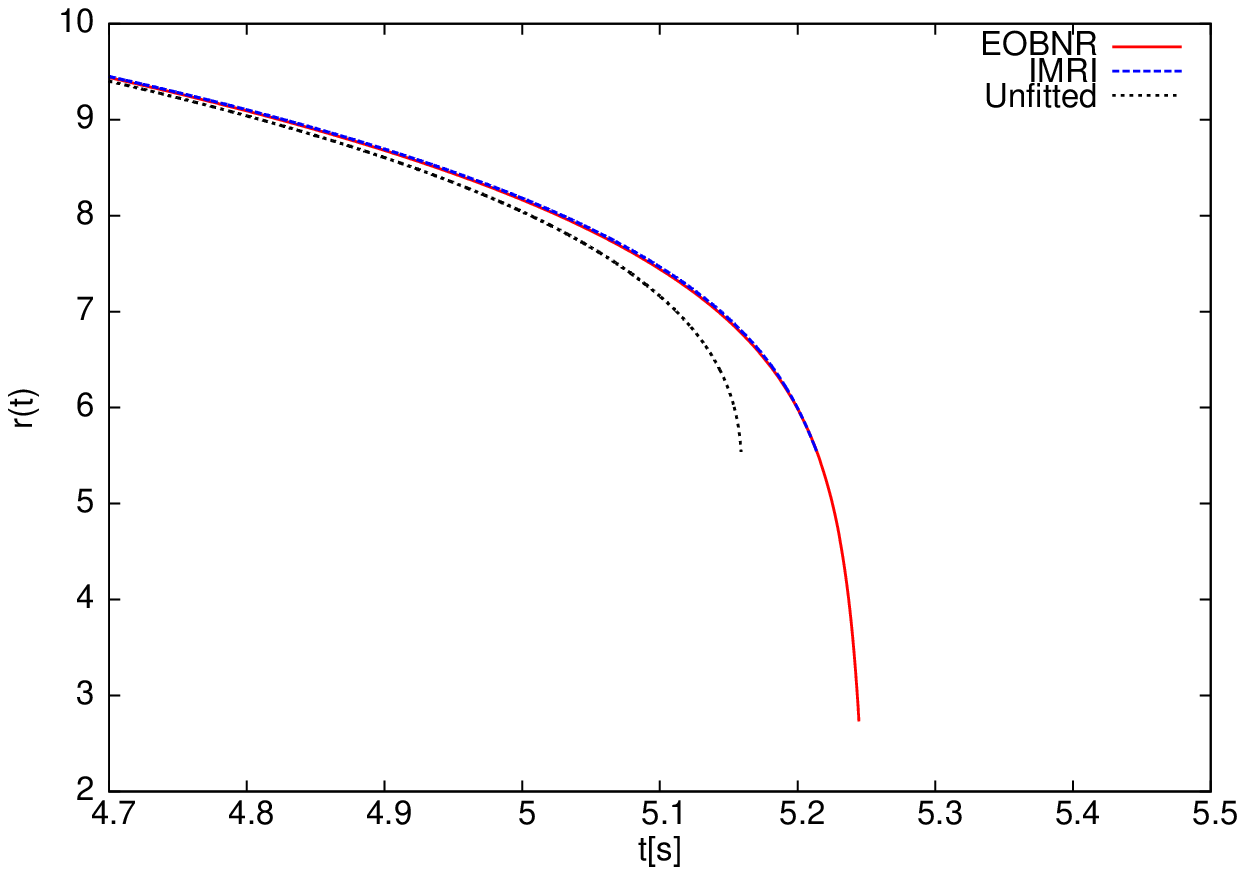}
}
\centerline{
\includegraphics[height=0.38\textwidth,  clip]{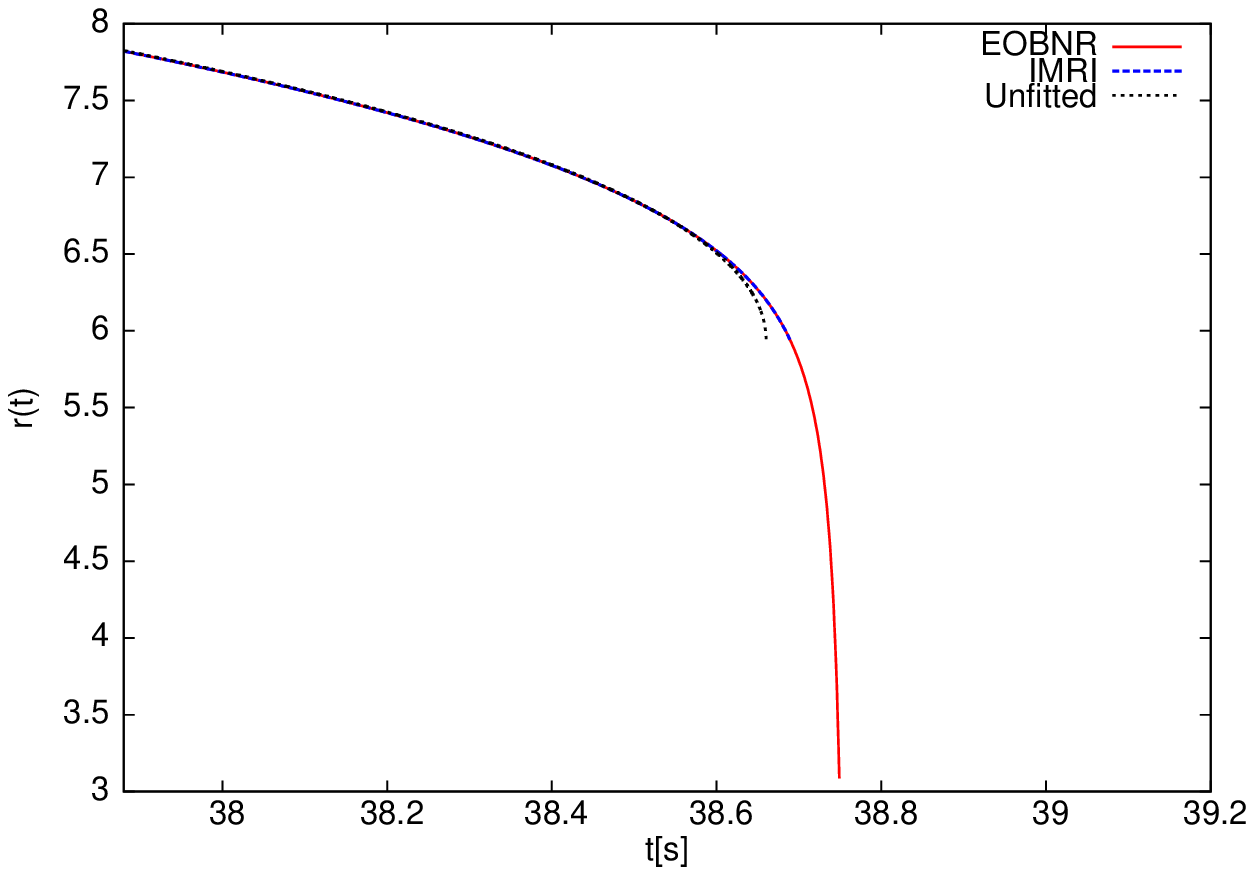}
\includegraphics[height=2.6in, width=3.8in,  clip]{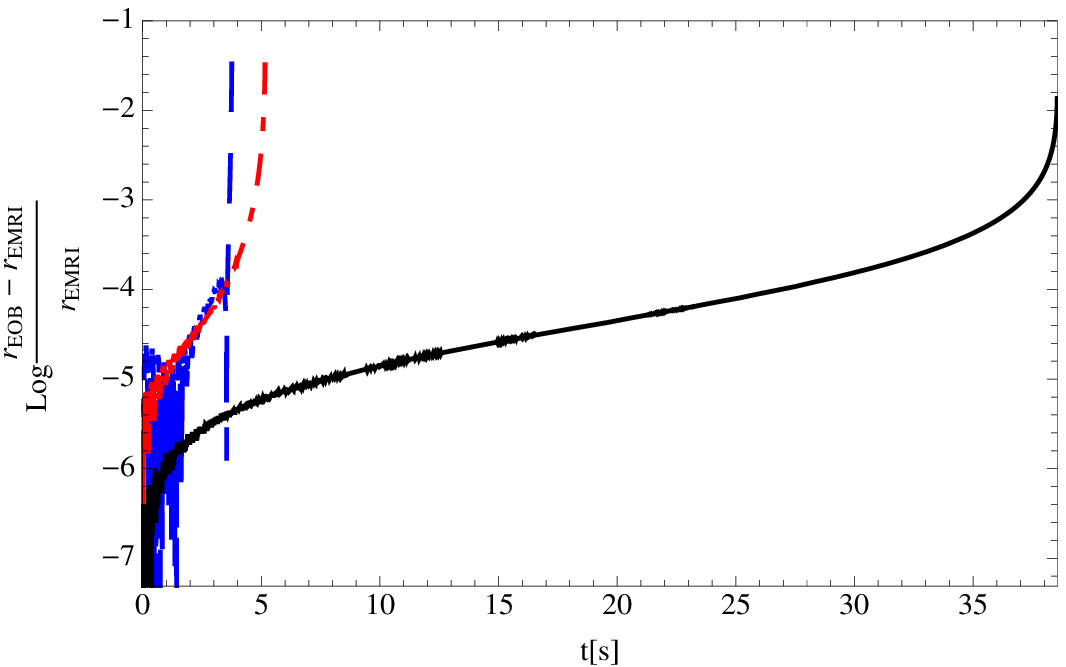}
}
\caption{The panels show the radial evolution using the EOBNRv2 model, the IMRI model described in the text, and a model (Unfitted) which is the same as the IMRI model described in text except for the fact that the coefficients used in Eq.~\eqref{zedsf} for the function  \(z_{\rm SF} (x)\) were derived using available self-force data from extreme-mass ratio calculations~\cite{barus}. The plots correspond to binaries of component masses \(17M_{\odot} + 100M_{\odot}\) (top-left panel), \(10M_{\odot} + 100M_{\odot}\) (top-right panel) and \(1M_{\odot} + 100M_{\odot}\) (bottom-left panel). The bottom-right panel shows that the orbital evolution predicted by a model that incorporates self-force corrections from extreme-mass-ratio inspirals (EMRIs) deviates from the orbital evolution predicted by the EOBNRv2 model at late inspiral. This discrepancy is more noticeable for systems with mass-ratios 17:100 (dashed blue) and 10:100 (dashed-dot red). Furthermore, for smaller mass-ratios, 1:100 (solid black), the discrepancy becomes comparatively smaller, as expected. }
\label{emimcomp}
\end{figure*}

In order to show the importance of increasing our knowledge of the self-force in the intermediate-mass-ratio regime, we present in Figure~\ref{emimcomp}  three different curves which describe the time evolution of three binary systems during late inspiral. These plots show that if we use a waveform model (unfitted) that includes: a) an accurate prescription for the orbital frequency \(\hat\Omega\); b) a prescription for the flux of angular momentum that incorporates the contribution from all dominant and subdominant \((\ell,m)\) modes; c) an invariant expression for the angular momentum using the object \(x=\hat\Omega^{2/3}\) (see  Eq.~\eqref{angmom}), and; d) available self-force corrections to constrain the coefficients \(b_i\)~\cite{barus}, then such a model would generate an inspiral trajectory that deviates from the expected orbital evolution, in particular near the ISCO.

Figure~\ref{emimcomp} also shows that our model actually predicts the expected orbital evolution all the way down to the ISCO, in agreement with the EOBNRv2 model, and requires a different set of coefficients  \(b_i\) in the redshift observable \(z_{\rm SF} (x)\), as compared with results in the extreme-mass-ratio limit quoted in~\cite{barus}. Notice that this is not only a modeling issue. It is an indication that implementing available self-force data into IMRI waveform models will not render the correct inspiral evolution, at the very  least for the cases we have considered. This is an important result of this paper. In order to substantiate this statement, in the following Section we will compute the value of the gauge invariant angular momentum \(L_z(x)\) at the ISCO within both the self-force formalism and our IMRI model. We will also show that the evolution of \(L_z(x)\) is consistent between both formalisms during early inspiral, but that the form of this object differs as we near the ISCO.  

One may also expect that for binaries with small mass-ratios, available self-force corrections may provide a fairly good description of the inspiral evolution. This is what we actually see in the bottom panel of Figure~\ref{emimcomp}. We will also show in the following Section that the evolution of \(L_z(x)\) for binaries with mass-ratio 1:100 is pretty consistent from early inspiral to the ISCO with EOBNRv2. This may not be surprising, since current self-force data have been obtained in the context of EMRIs and the EOBNRv2 has been calibrated so as to reproduce the dynamical evolution of binary black holes of small mass-ratio~\cite{resu,raci,nic,nic1}. This exercise then suggests that it may be necessary to go beyond first-order conservative corrections to reproduce accurately the expected dynamical evolution of intermediate-mass ratio systems with \(\eta \sim 10^{-2}-10^{-1}\).

To carry out the analysis described above in the following Section, we will start by using the EOBNRv2 model to compute the orbital frequency ISCO shift for binaries of mass ratio 1:1, 1:2, 1:3, 1:4, 1:5, 1:6, 1:10 and 1:100, along with perturbative results in the context of extreme-mass-ratio inspirals. We will use this expression for the orbital frequency ISCO shift  to evaluate the gauge-invariant object \(x\)  and then compute the value of the angular momentum at the ISCO using Eq.~\eqref{angmom}.  We should also acknowledge the fact that EOB has not yet been calibrated using NR simulations of systems with mass-ratio 1:100. It is expected that EOBNRv2 provides a fair description of the dynamical evolution of these types of sources,  and we show in the following section that both EOBNRv2 and perturbative calculations provide a very consistent modeling of the gauge-invariant angular momentum \(\hat{L}_{z}\) from early inspiral to the ISCO for binaries with mass-ratio 1:100. However, this study should be compared to accurate NR simulations of mass-ratio 1:100~\cite{carlos} once these are generated with several gravitational waveform cycles before merger.

\clearpage

\section{ISCO shift: connecting the extreme, intermediate and comparable-mass ratio regimes}
\label{s3}

In the previous Sections we have mentioned that during the inspiral of a stellar mass compact object of mass \(m_2\) into a supermassive BH of mass \(m_1\), the radiative part of the self-force drives the inspiral evolution of the small object, whereas its conservative part has a cumulative effect on the orbital phase evolution \cite{SFB}. These two effects have been considered in the development of EMRI waveform templates  \cite{amos, cutler, gairles, kludge, improved, lisacapt, seoane}. 

We shall now consider a novel effect that was explored by Barack \& Sago~\cite{inner}. They have shown that the self-force also introduces shifts in the innermost stable circular orbit radius and frequency. For a test-mass particle, these two quantities are given by

\begin{equation}
r_{\rm ISCO} = 6m_1,  \qquad m_1\Omega_{\rm ISCO} = \frac{1}{6\sqrt{6}},
\label{tmpt}
\end{equation}

\noindent whereas, for finite \(\eta\), in the Lorenz gauge, these two quantities take the form~\cite{inner} 

\begin{equation}
\Delta r_{\rm ISCO} = -3.269(\pm3\times10^{-3}) m_2,  \qquad \frac{\Delta\Omega_{\rm ISCO}}{\Omega_{\rm ISCO}} = 0.4870(\pm6\times10^{-4})\frac{m_2}{m_1}.
\label{shift}
\end{equation}

Since Barack \& Sago carried out these calculations in Lorenz gauge, it was necessary to translate these results into coordinates that are commonly used for GW observations, i.e., asymptotically flat coordinates.  This exercise has been done for the orbital frequency, which is a gauge invariant object, and hence can be compared to results obtained in alternative formalisms, such as PN theory or the EOB approach.  In \cite{baracknewphi} and \cite{damsh}, the authors derive the `renormalization' factor that translates results of Lorenz-gauge calculations into  physical units. Applying this renormalization technique, one finds that the ISCO frequency is given by  

\begin{equation}
M\Omega_{\rm ISCO} = \frac{1}{6\sqrt{6}}\left(1+1.2512\eta + {\cal{O}}(\eta^2)\right),
\label{bsshift}
\end{equation}

\noindent with \(M=m_1+m_2\). This prediction can be compared with PN--based ISCO calculations at different orders of accuracy \cite{favata}

\begin{eqnarray}
M\Omega^{2{\rm PN}}_{\rm ISCO} &=&\frac{1}{6\sqrt{6}}\left(1+\frac{7}{12}\eta + {\cal{O}}(\eta^2)\right), \\\nonumber
M\Omega^{3{\rm PN}}_{\rm ISCO} &=&\frac{1}{6\sqrt{6}}\left(1+ \left(\frac{565}{288} - \frac{41}{768}\pi^2\right)\eta + {\cal{O}}(\eta^2)\right). \\\nonumber
\label{pnshift}
\end{eqnarray}

\noindent The EOB approach has also been used to describe the orbital frequency shift at ISCO. Damour suggested in \cite{damsh} that a fit for the ISCO orbital frequency shift that incorporates results from the gravitational self-force and NR simulations may be a quadratic polynomial in \(\eta\) of the form \cite{damsh} 

\begin{equation}
M\Omega^{\rm Damour}_{\rm ISCO} = \frac{1}{6\sqrt{6}}\left(1+1.25\eta + 1.87\eta^2\right).
\label{damshift}
\end{equation}

\noindent In this paper we build up on this analysis and update this estimate using results from the self-force program, and making use of the EOBNRv2 model, which has been calibrated to NR simulations~\cite{buho}. The prescription for the orbital frequency ISCO shift that we propose below reproduces accurately the results predicted by the self-force program for EMRIs, and also reproduces the best data currently available for intermediate and comparable-mass systems.  To compute the ISCO orbital frequency shift within the EOB formalism, we use the equation derived in \cite{damsh}, namely,

\begin{equation}
2A(u)A'(u) + 4u\left(A'(u)\right)^2 - 2uA(u)A''(u)=0,
\label{damisco}
\end{equation}

\noindent where \(u=1/r\) and \('\) stands for \(d/du\). This ISCO condition can be rewritten in terms of the radial coordinate as \cite{favata}

\begin{equation}
rA(r)A''(r) -2r\left(A'(r)\right)^2 +3A(r)A'(r)=0,
\label{iscoeq}
\end{equation}

\noindent where \('=d/dr\). We shall use the metric coefficient \(A(r)\) quoted in \cite{buho}, which includes the Pad\'e expression for \(A(r)\) at 5PN order.  Having obtained the value for \(r_{\rm  isco}\) using Eq.~\eqref{iscoeq}, we evaluate the angular orbital frequency  \(M\Omega_{\rm ISCO}\) at  this fiducial value using Eq. (10b) of \cite{buho} with \(p_r =0\). 

Using a variety of events, including extreme, \(\eta\sim 10^{-5}\), and intermediate, \(\eta\sim 10^{-2}-10^{-1}\), mass ratio inspirals, we derive a quadratic polynomial fit  in \(\eta\) for the ISCO orbital frequency shift for these types of events, namely

\begin{equation}
M\Omega^{\rm fit}_{\rm ISCO} = \frac{1}{6\sqrt{6}}\left(1+1.05786\eta + 2.12991\eta ^2\right).
\label{newshift}
\end{equation}

\noindent We found that at second order in \(\eta\), this numerical fit does not reproduce accurately the orbital frequency ISCO shift for small-mass ratios~\cite{inner}. We can fix these problems using a prescription of the form

\begin{equation}
M\Omega^{\rm fit}_{\rm ISCO} = \frac{1}{6\sqrt{6}}\left(1 + 1.2512\eta - 0.0553751\eta^2 + 5.78557\eta^3\right).
\label{comisco}
\end{equation}

\begin{figure*}[ht]
\centerline{
\includegraphics[height=0.44\textwidth,  clip]{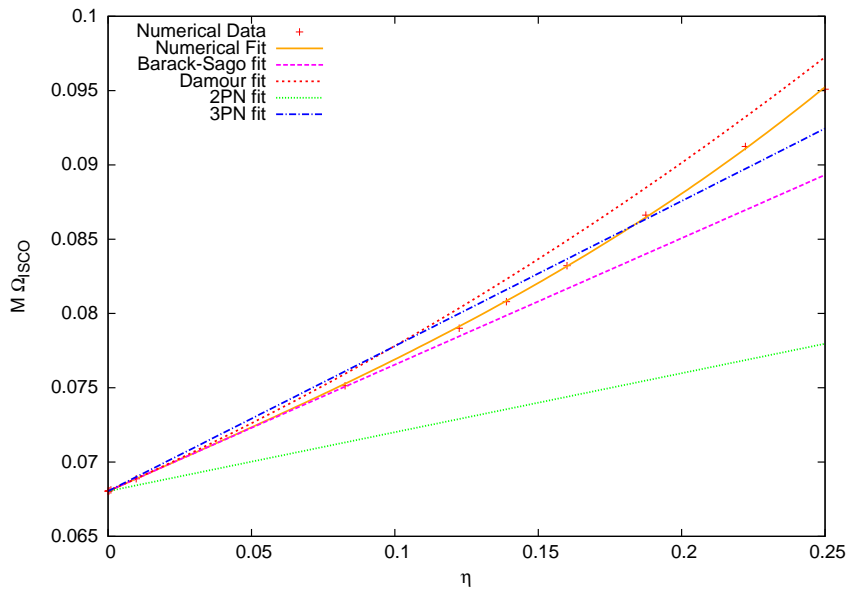}
}
\caption{ISCO shift using various approximations as described in the text. The `Numerical Data' has been obtained using calculations in the extreme-mass ratio regime~\cite{inner} and the EOBNRv2. The prescription that encapsulates results from the extreme, intermediate and comparable mass-ratio regime is labeled as `Numerical Fit' and is given by Eq.~\eqref{comisco} in the main text.  }
\label{iscoshift}
\end{figure*}

\noindent At this level of accuracy we exactly reproduce the prediction for EMRIs for small \(\eta\), as well as the most-up-to-date results for binaries modeled using the EOBNRv2 scheme. We compare the range of applicability of this numerical expression, along the other various approximations mentioned above,  in Figure~\ref{iscoshift}.

\begin{figure*}[ht]
\centerline{
\includegraphics[height=0.38\textwidth,  clip]{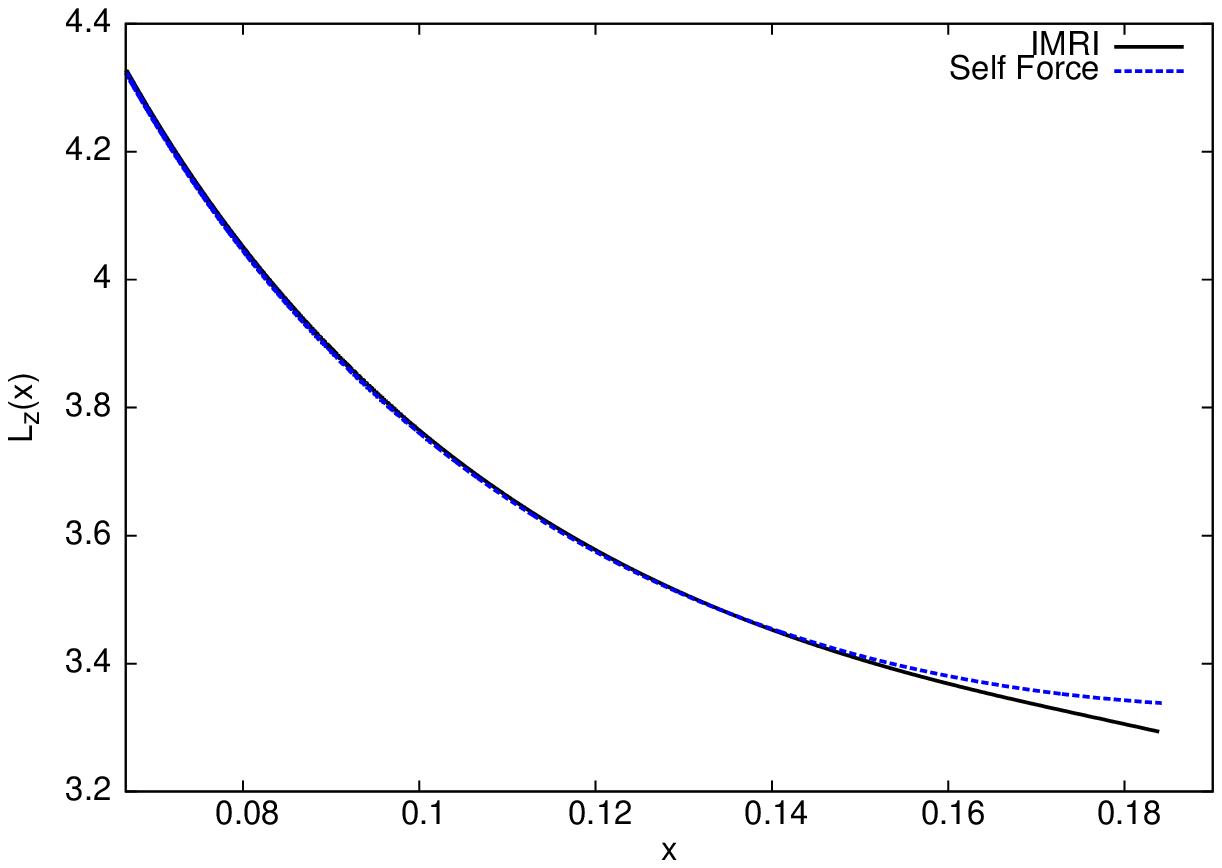}
\includegraphics[height=0.38\textwidth,  clip]{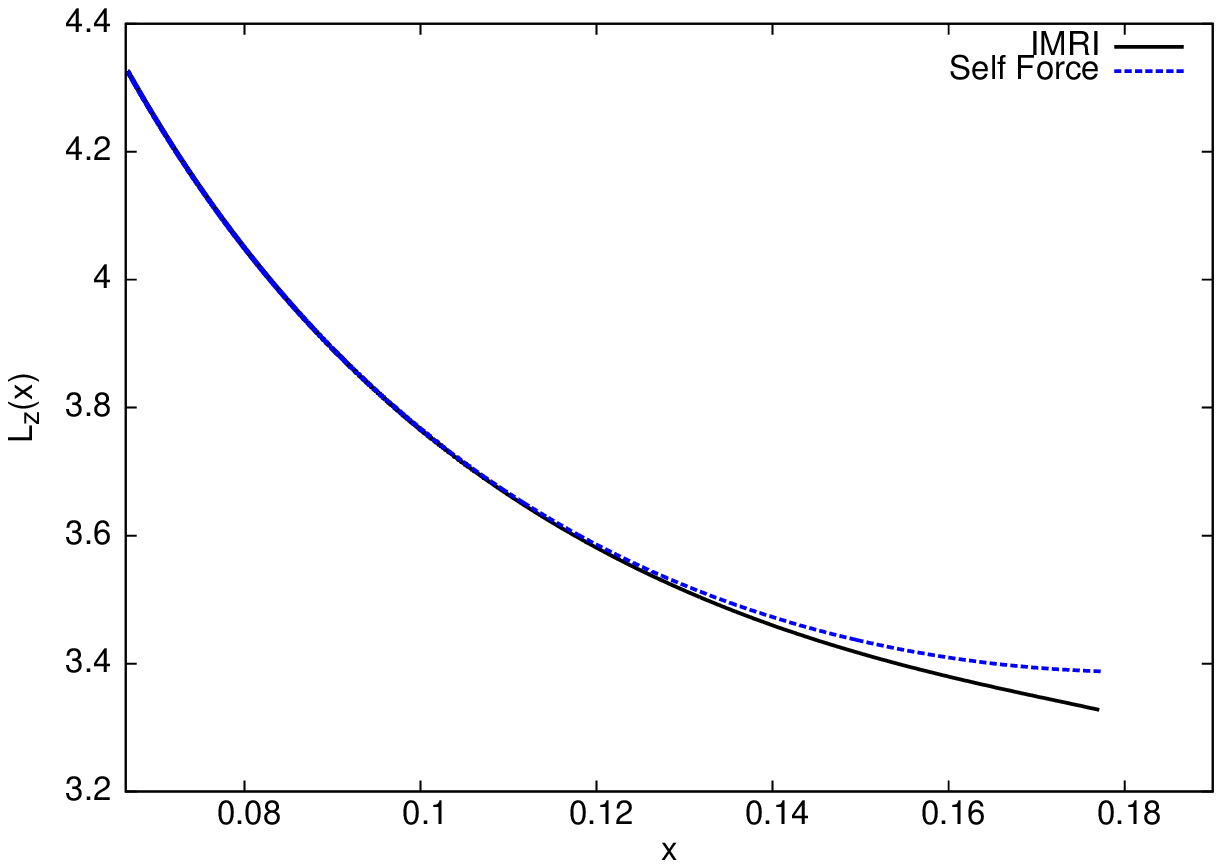}
}
\centerline{
\includegraphics[height=0.38\textwidth,  clip]{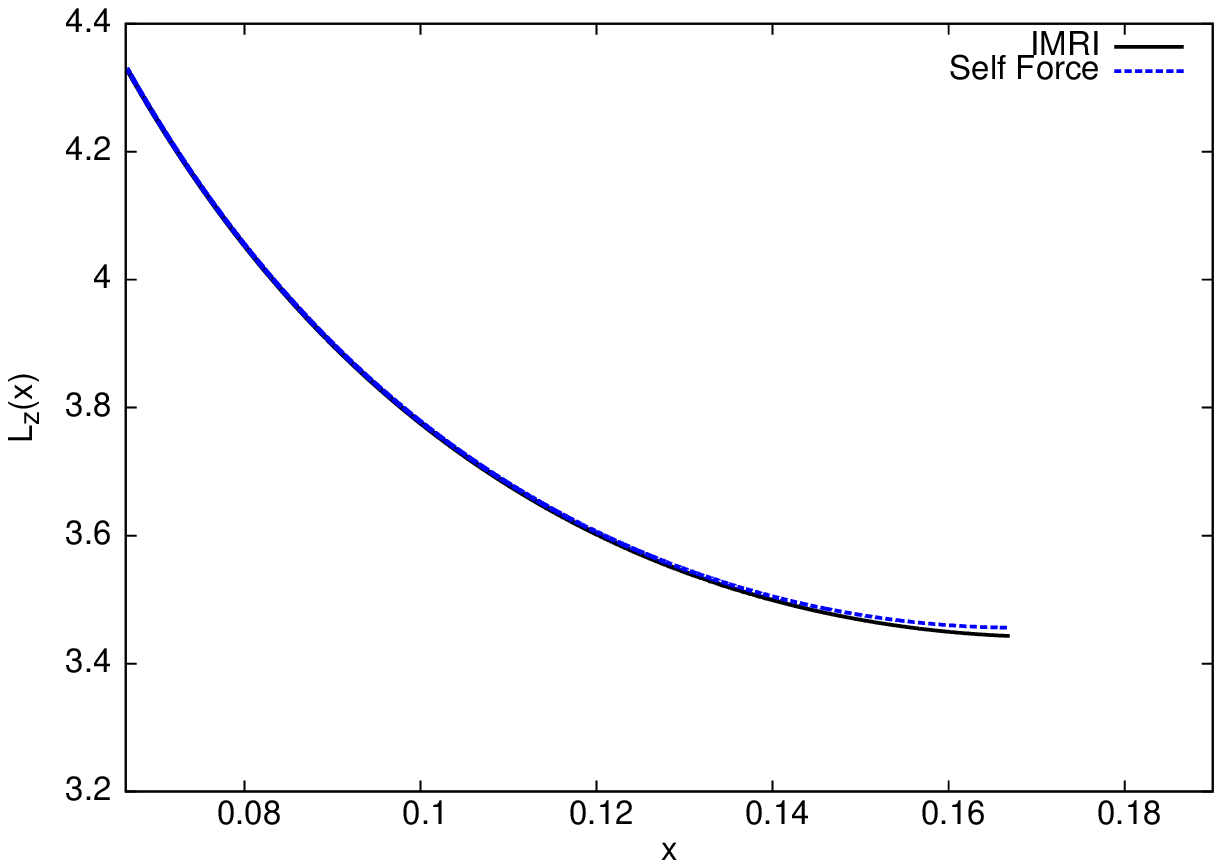}
\includegraphics[height=2.6in, width=3.8in,  clip]{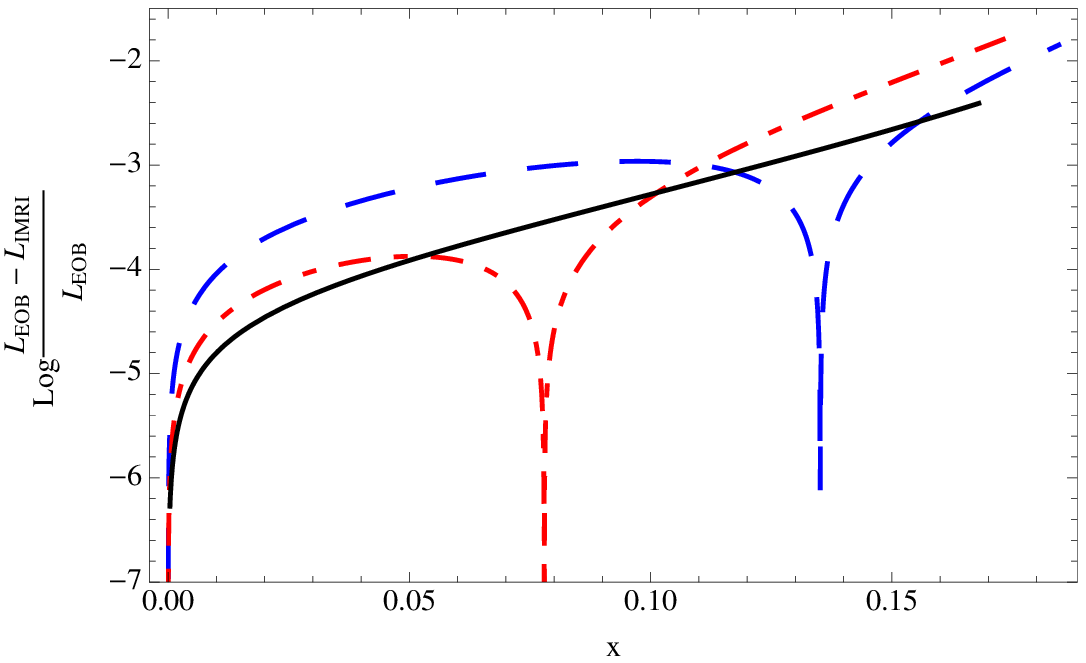}
}
\caption{The panels show the invariant angular momentum \(L_{z}(x)\), with \(x=\left(M\Omega\right)^{2/3}\), from early inspiral to the ISCO using two different prescriptions. The `IMRI' prescription reproduces accurately the expected inspiral evolution, as compared with results from the EOBNRv2 model ---which has been calibrated to NR simulations. The `Self-Force' prescription encapsulates self-force corrections derived in the context of EMRIs~\cite{sago,barus}. Note that for the three binary systems studied,  \(17M_{\odot} + 100M_{\odot}\) (top-left panel),  \(10M_{\odot} + 100M_{\odot}\) (top-right panel), and  \(1M_{\odot} + 100M_{\odot}\) (bottom-left panel), the `Self Force' prescription for the angular momentum presents a deviation from its expected value that becomes more noticeable near the ISCO. The bottom-right panel shows the fractional accuracy between the two different prescriptions for the angular momentum for the systems 17:100 (dashed blue), 10:100 (dashed-dot red) and 1:100 (solid black). Notice also that, as expected, for binaries with small mass-ratio the `Self Force' and `IMRI' prescriptions are fairly consistent all the way down to the ISCO.  As in the previous plots, the spikes in the bottom-right panel are due to numerical artifacts.}
\label{lzfull}
\end{figure*}

With our prescription for \(M\Omega_{\rm ISCO}\), we can evaluate the value of the gauge-invariant angular momentum flux at the ISCO. To do so, we compute the value of the angular momentum, see Eq.~\eqref{angmom}, using available self-force data (SF Fit), i.e., we use Eq.~\eqref{bsshift} to compute the shift in the orbital frequency at the ISCO, and then use Eq.~\eqref{angmom} in conjunction with the values for the \(b_i\) coefficients of Eq.~\eqref{zedsf} quoted in~\cite{barus}, i.e., self-force corrections derived in the extreme-mass-ratio limit. We also present results for an `incomplete model', in which we use Eq.~\eqref{comisco} to evaluate the value of the orbital frequency at the ISCO, and then use Eq.~\eqref{angmom} with the set of \(b_i\) coefficients quoted in~\cite{barus}. Finally, the IMRI model encodes all the results derived in this paper, namely, we use the prescription for the shift of the orbital frequency at the ISCO in Eq.~\eqref{comisco}, and the prescription for the angular momentum (Eq.~\eqref{angmom}) using the corrections quoted in Table~\ref{zedfunccoef}.  Table~\ref{lzcomp} shows that, in accord with Figure~\ref{emimcomp}, for binaries with symmetric mass-ratio \(\eta \sim 0.01\), the value of the angular momentum evaluated at ISCO is fairly consistent between the two models. However,  the predicted value of the angular momentum at ISCO becomes more discrepant for binaries with   \(\eta \sim 0.1\). This then suggests that the evolution of intermediate-mass ratio inspirals cannot be fully captured by using self-force calculations from extreme-mass ratio inspirals. This can be better visualized in Figure~\ref{lzfull} where we show the angular momentum during inspiral all the way down to the ISCO using two formalism, namely, the IMRI prescription and a model that includes available self-force corrections, which is labeled as `Self Force'.

\begin{table}[thb]
\begin{tabular}{|c|c|c|c|}
\hline\multicolumn{1}{|c|}{}&\multicolumn{3}{c|}{\(L_z(x_{\rm ISCO} )\)}\\\cline{2-4}
\multicolumn{1}{|c|}{\(\eta\)}&SF Fit&Incomplete&IMRI \\\cline{1-4}
$\frac{1700}{13689}$&3.3389 &3.3522&3.2918 \\[1ex]\cline{1-4}
$\frac{10}{121}$& 3.3876&3.3878&3.3254\\ [1ex]\cline{1-4}
$\frac{100}{10201}$& 3.4561&3.4561&3.4429\\[1ex]\hline
\end{tabular}
\caption{ The Table shows the value of the angular momentum \(L_z\) as a function of the gauge invariant object \(x=\left(M\Omega \right)^{2/3}\)  evaluated at the ISCO radius (see Eq.~\eqref{angmom}). The SF (self-force) values are computed using the prescription given in Eq.~\eqref{bsshift} to compute the orbital frequency shift, and Eq.~\eqref{angmom} with the coefficients \(b_i\) quoted in \cite{barus}, i.e., evaluated in the context of extreme-mass-ratio inspirals. Incomplete stands for a prescription in which we use the prescription for the orbital frequency given by Eq.~\eqref{comisco}, and the angular momentum prescription given by  Eq.~\eqref{angmom} with the coefficients \(b_i\) quoted in \cite{barus}. The   \(b_i\) values of the IMRI model are obtained using a model that reproduces the dynamical evolution of intermediate-mass-ratio inspirals, as compared with the EOBNRv2 model, and for which we have derived a new prescription for the red-shift observable \(z_{\rm SF} (x)\) which actually reproduces the features of true inspirals.}
\label{lzcomp}
\end{table}

Table~\ref{lzcomp} and Figures~\ref{emimcomp}, \ref{lzfull} show that as the mass-ratio increases, the discrepancy between a model that incorporates available self-force corrections~\cite{sago} and one that has been calibrated to NR simulations becomes more pronounced. For events with mass-ratio 1:10 this difference looks slightly larger than for those of mass-ratio 17:100. Part of the reason for this behavior may be the fact that the model we have used to perform this analysis, EOBNRv2, provides the best prescription currently available for the inspiral evolution of sources with mass-ratios 1:1-1:6. It is expected that the model provides a good description of the inspiral evolution of sources with mass-ratio 1:10, and hence we have extended its realm of applicability to shed light on the form that the self-force is expected to have so as to reproduce the inspiral evolution of these type of sources. Thus, a conservative conclusion we can draw at this stage is that, using the best waveform model currently available in the literature, we have shown that  one may not be able to accurately model the dynamics of sources with mass-ratio 1:10 using available self-force calculations.  This trend is also present   in the case of events with mass-ratio 17:100 and the results in this case are more conclusive. We have shown that a waveform model that includes available self-force corrections will not be able to capture faithfully the inspiral evolution of these GW sources. These results suggest that one may need to include higher-order corrections in the self-force to capture the behavior of true inspirals in the intermediate-mass-ratio regime. Finally, our work is also a consistency check on the internal structure of the EOB model, since we have shown that the EOBNRv2 renders a good prescription for the inspiral evolution via the flux of angular momentum and that the angular momentum prescription is consistent with results obtained from perturbative calculations. 
\clearpage

 \section{Conclusions}
 \label{s4}
 
 In this paper we have developed a code to reproduce the analysis presented in \cite{buho}. Using this code we generated the inspiral evolution of three different systems to perform an exploratory study of the importance of including accurate self-force corrections in search templates that aim to detect non-spinning intermediate-mass ratio inspirals. The choice of the systems to perform this study reflects the knowledge we have at present on the dynamical evolution of non-spinning BH binaries (\(17M_{\odot} + 100M_{\odot}\)), what we want to know (\(10 M_{\odot} + 100M_{\odot}\)), and a special system that  provides reassurance that the dynamics of intermediate-mass-ratio inspirals with typical mass-ratios \(\eta\sim 0.01\) can be captured using perturbative theory, as shown in \cite{carlos,ulr}. 
 
 The EOBNRv2 model we have used as a benchmark to carry out this study has the advantage of encoding the best information currently available of non-spinning BH binaries, is capable of reproducing the expected dynamics in the test-mass particle limit, and also includes corrections taken from self-force corrections in the extreme-mass-ratio limit. We have confirmed these statements for systems with  \(\eta\sim 0.01\) by showing  that the EOBNRv2 does predict the expected form of the angular momentum flux, as compared to Teukolsky data, as well as the evolution  of the gauge-invariant angular momentum, as compared to self-force data. 
 
In order to explore the form of the self-force in the intermediate-mass-ratio regime, we developed an IMRI model that reproduces the inspiral evolution predicted by the EOBNRv2 model, and which enable us to explore the form that the self-force should have in this mass-ratio regime so as to reproduce the binary's dynamical evolution as predicted by the best available interface to NR simulations. We have found that for systems with component masses  \(17M_{\odot} + 100M_{\odot}\), available self-force corrections do not accurately reproduce the inspiral evolution. We have shown that there is a clear deviation from the true inspiral trajectory near the ISCO. We have also explored this issue in greater detail by showing that the gauge-invariant angular momentum does deviate from the current self-force prediction near the ISCO. When we extend the realm of applicability of the EOBNRv2 to binaries of mass-ratio 1:10, we observe a similar behaviour.  This exploratory study then suggests that it may be necessary to extend conservative corrections beyond the linear order to accurately capture the true inspiral evolution for sources with intermediate-mass-ratio. Furthermore, once NR simulations of systems with mass-ratios 1:10, 1:15 and 1:100 have reached enough resolution near merger, we will be in a good position to calibrate the EOB model so as to reproduce the true dynamical evolution for these type of sources \cite{carlos, carlosI,carlosII}. Such a model will enable us to further probe the parameter space to develop IMRI models, as the one proposed in this paper, that capture the features of true inspirals at a very inexpensive computational cost. 

We have also used our EOBNRv2 code to derive a prescription for the orbital frequency ISCO shift that encapsulates results from the extreme, intermediate and comparable-mass ratio regimes. Our prescription is the first in the literature that reproduces the self-force result in the appropriate limit.   We have made use of this new prescription to estimate the value of the angular momentum at the ISCO using our IMRI prescription and available self-force data. We have found a clear discrepancy for systems with mass-ratios \(\eta \sim 0.1\), but have confirmed that for systems with small mass ratios \(\eta\sim 0.01\) both predictions are fairly consistent. 

This study has also shed light on  a pressing issue that needs to be addressed before aLIGO begins observations, namely, at present we use templates  in searches for GW sources  whose actual dynamics are currently unknown. For instance, searches for BH binaries with total mass  \(25M_{\odot} - 200M_{\odot}\), and individual masses from \(3M_{\odot}\) to \(99M_{\odot}\) have been carried out, but not a single model has been calibrated using high-resolution NR simulations for mass-ratios smaller than 1:6. Hence, it is important that numerical relativists and search template developers interact more closely so as to run NR simulations which cover regions in parameter space where future GW detectors may detect GW sources. This meaningful interaction will be important from a theoretical perspective, as we will be able to further our knowledge of the self-force, and from a data analysis perspective, as we will be in a stronger position to develop accurate search templates for use in data analysis. 

The approach we have outlined in this paper is the initial step to construct inspiral-merger-ringdown IMRI waveforms. Having derived a consistent prescription for the inspiral evolution, we can use a similar approach to that described in~\cite{ori,firstpaper} to include merger and ring-down in a physically consistent way. Developing complete IMRI waveforms may be useful for aLIGO data analysis, as they will provide the accuracy of more complex waveform models at an inexpensive computational cost.  
  
 \section*{Acknowledgments}
We thanks the Gravitational Wave Group at Syracuse University for useful discussions and Stefan Ballmer, Ryan Fisher, Ian Harry, Alex Nitz  and Peter Saulson for useful interactions and criticisms to this work. We would like to thank Evan Ochsner for carefully reading this manuscript and for providing valuable criticism to improve the description of the construction of our IMRI waveform. We also thank Alessandra Buonanno and Yi Pan for useful comments and feedback. This work was supported by NSF grant PHY-0847611. DB and EH would also like to thank the Research Corporation for Science Advancement for support. The simulations described in this paper were performed using the  Syracuse University Gravitation and Relativity  (SUGAR) cluster,  which is supported by NSF grants PHY-1040231, PHY-1104371 and PHY-0600953.

\section{Appendix}

In this Section we show that the approach outlined in the main body of the paper can still be used to model equal-mass (EM) binaries with great accuracy at an inexpensive computational cost. To do so we consider a system with total mass \(20 M_{\odot}\). Following the method described in Section~\ref{s2}, we start by deriving accurate prescriptions for the orbital frequency \(\Omega_{\rm ansatz}\) and angular momentum flux \(\dot{L}_z\). Thereafter we derive the corrections that should be implemented in the gauge-invariant expression for the angular momentum (see Eq.~\eqref{angmom}) to accurately reproduce the inspiral evolution predicted by the EOBNRv2 model. 

\begin{figure*}[ht]
\centerline{
\includegraphics[height=2.6in, width=3.8in,  clip]{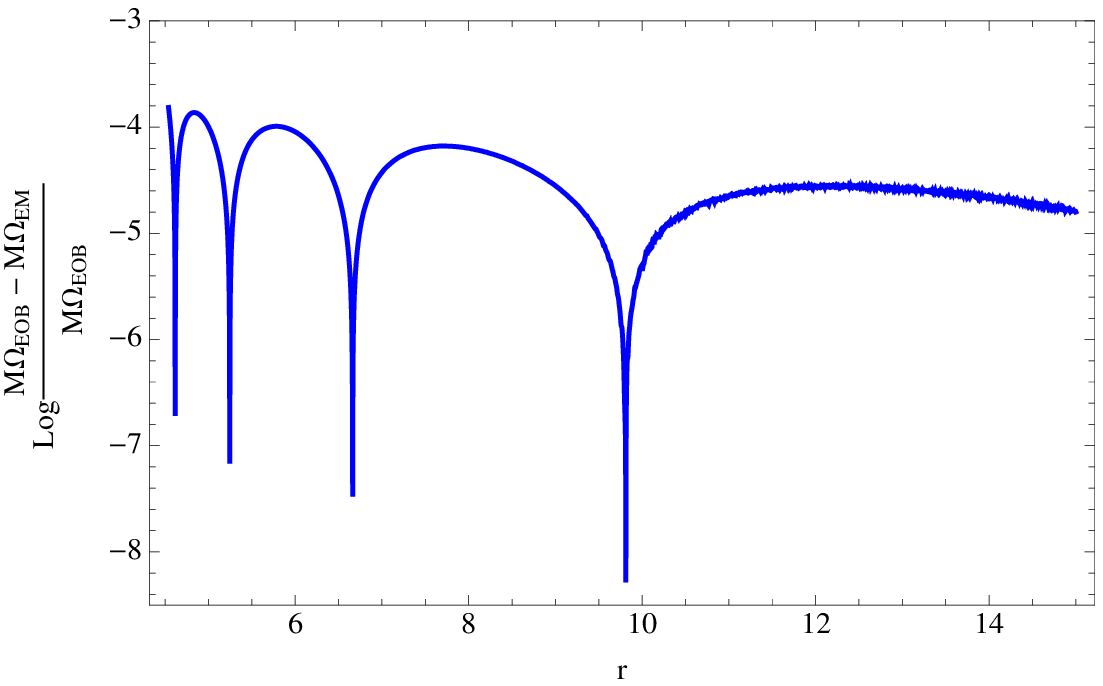}
\includegraphics[height=2.6in, width=3.8in,  clip]{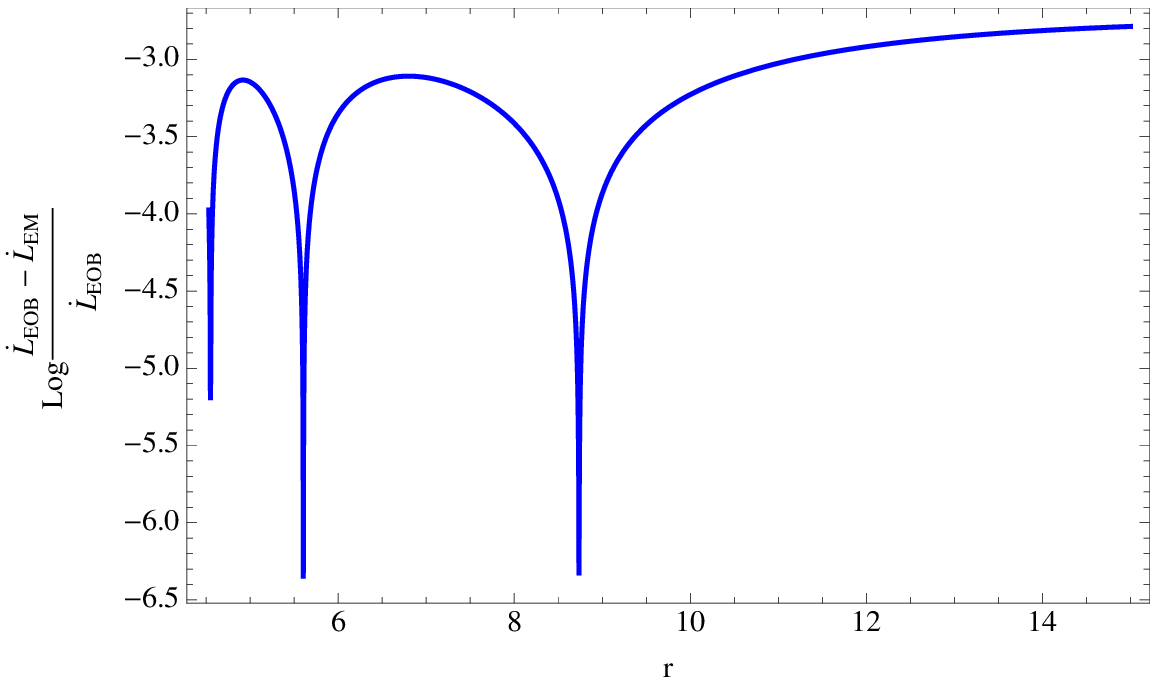}
}
\centerline{
\includegraphics[height=2.6in, width=3.8in,  clip]{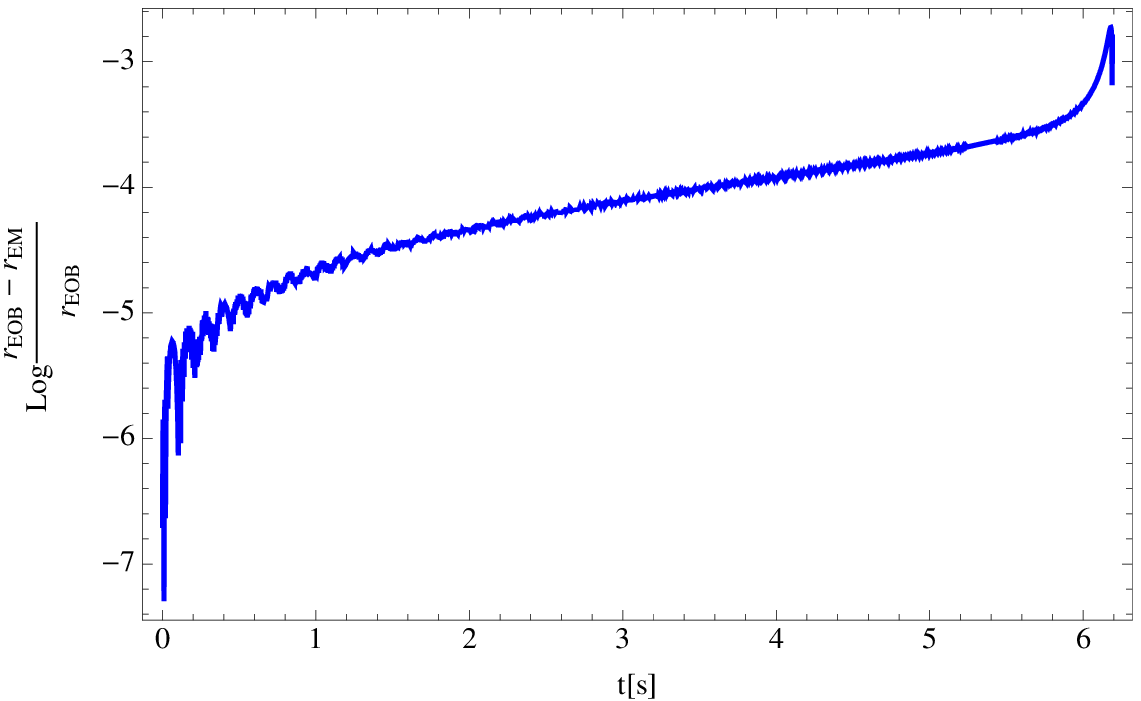}
\includegraphics[height=2.6in, width=3.8in,  clip]{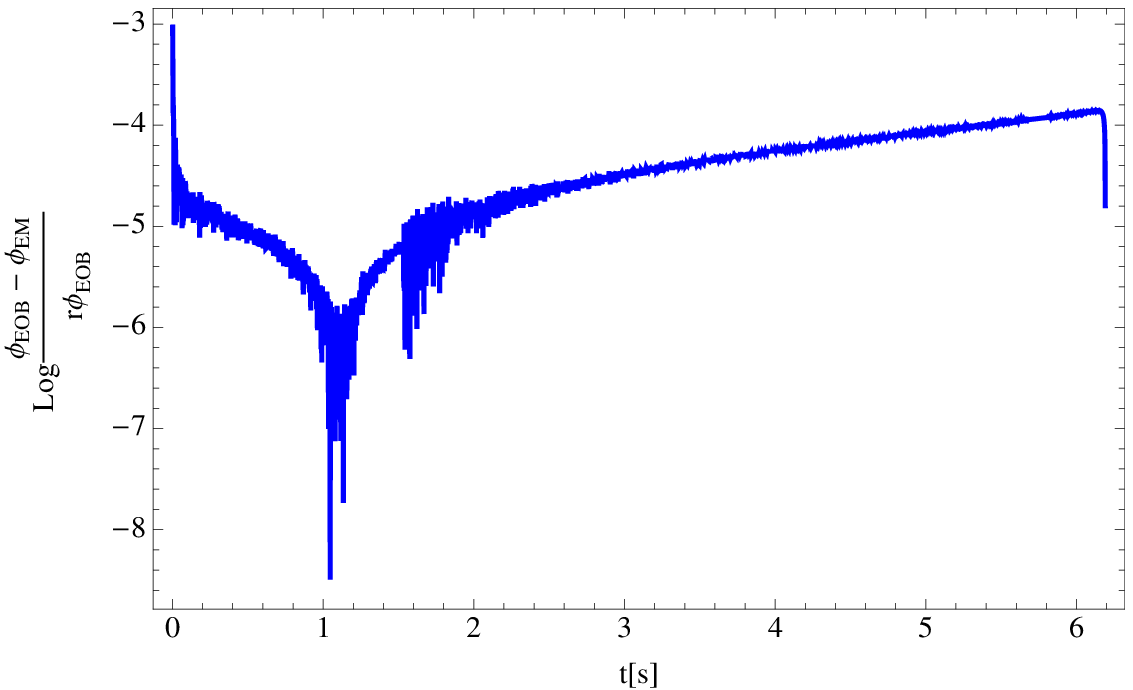}
}
\centerline{
\includegraphics[height=2.6in, width=3.8in,  clip]{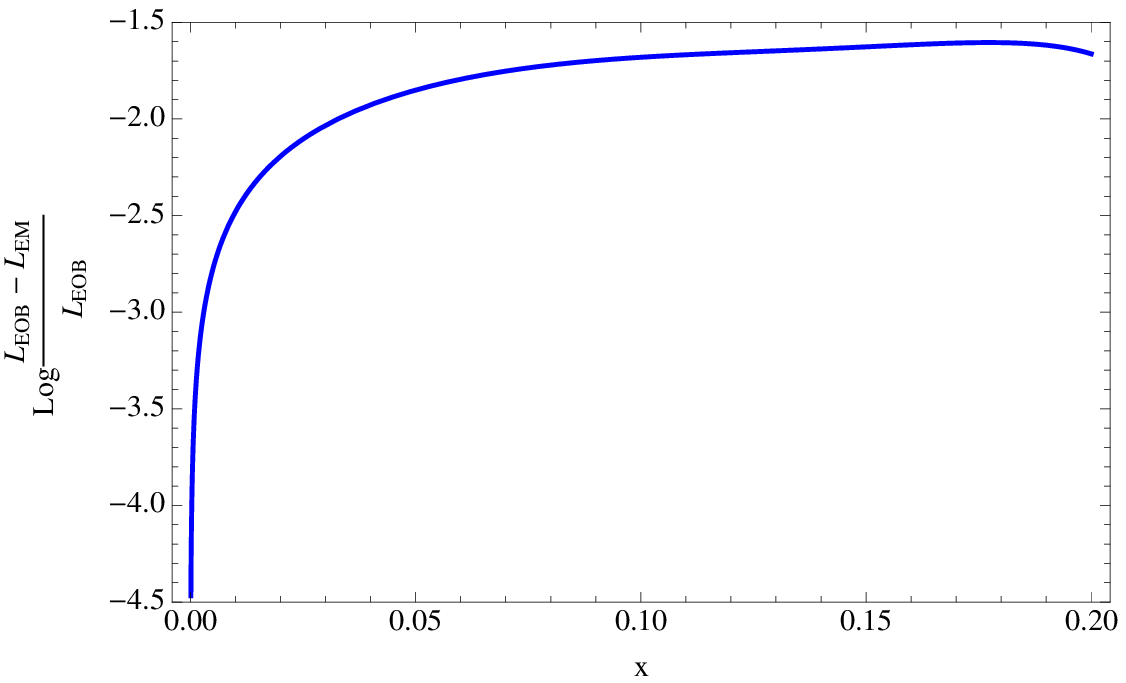}
}
\caption{The panels show the accuracy with which our model can reproduce the dynamics of an equal mass (EM) binary system with masses  ( \(10M_{\odot} + 10M_{\odot}\)), as compared with the EOBNRv2 model. The top panels show that our model can reproduce the orbital angular evolution and the flux of angular momentum predicted by the EOBNRv2 model point to point with an accuracy better than one part in a thousand. The middle panels show that our model can reproduce with a similar accuracy the radial and azimuthal evolution of a   \(10M_{\odot} + 10M_{\odot}\) binary system. The bottom panel shows that the expression for the gauge-invariant angular momentum \(L_{z}(x)\) for an EM mass deviates clearly from the EMRI prescription even before nearing the ISCO. }
\label{emcase}
\end{figure*}

The prescriptions for the orbital frequency, angular momentum flux and angular momentum which reproduce the inspiral evolution for the \(10M_{\odot} + 10M_{\odot}\) binary system shown in Figure~\ref{emcase} require the following set of coefficients (compare Tables~\ref{coef}, \ref{Lzdotcoef}, \ref{zedfunccoef})

\begin{eqnarray}
\Omega(u)_{\rm ansatz} &:& a_1= -5.644, \quad a_2 = 11.003, \quad a_3 = -5.390,\quad a_4 = 8.794, \quad a_5 = 0.458\\\nonumber
\dot{L}_z &:& c^{1}_{2.5} = -90.566, \quad c^2_3  =304.941, \quad c^3_{3.5}= -339.500\\\nonumber
z_{\rm SF} &:&  b_1 = -5.803,\quad  b_2= 14.171, \quad b_3 = -7.074, \quad b_4 = 22.810, \quad b_5 =-25.580\\\nonumber
\label{ecmcoe}
\end{eqnarray}

Having shown that the approach outlined in the paper is still effective to model EM binaries, we consider worthwhile developing a model that encapsulates the physics of binaries of comparable and intermediate-mass-ratio. In order to extend our IMRI model into the comparable mass-ratio regime, we require to implement some improvements in the model.  First and foremost, we require expressions for the orbital frequency evolution (see Eq.~\eqref{ompres}) and angular momentum (see Eq.~\eqref{angmom}) which incorporate conservative corrections at second-order in mass ratio \(\eta\). We can draw this conclusion by comparing the bottom-right panel of Figure~\ref{lzfull} with the bottom panel of Figure~\ref{emcase}. The former plot shows that for IMRIs the prescription for the angular momentum  deviates from the self-force prediction near the ISCO. However, the latter plot shows that in the EM case the prescription for the angular momentum deviates considerably from the self-force prediction even during the inspiral evolution, far away from the ISCO. Because the prescription we have used for the angular momentum in both cases includes corrections at liner order in mass-ratio, this suggest that to derive a model that unifies both regimes we require  an expression for the angular momentum which includes second-order conservative corrections. We can follow a similar strategy to encapsulate in a single prescription the radiative piece of the self-force for comparable and intermediate mass-ratio systems by deriving second-order radiative corrections in the prescription for the angular momentum flux. 

Having derived expressions for \(\Omega_{\rm ansatz}\), \(\dot{L}_z\) and \(L_{z}(x)\) which include corrections at second-order in mass-ratio, we will be able to tune the coefficients of these objects and derive numerical fits for the coefficients in terms of the mass-ratio \(\eta\) of the system. This approach will provide a unified description for the inspiral evolution of IMRIs and comparable-mass systems. We will pursue these studies in the future.

\clearpage

\bibliography{references}

\end{document}